\documentclass[journal,onecolumn,12pt,twoside,letterpaper,draftclsnofoot]{IEEEtran}
\pdfoutput=1
\usepackage{booktabs}
\usepackage{subfigure}
\usepackage{array}
\usepackage{amsmath}
\usepackage{amssymb}
\usepackage{graphicx}
\usepackage{cite}
\usepackage[usenames,dvipsnames,svgnames,table]{xcolor}
\usepackage{comment}
\usepackage{tablefootnote}
\allowdisplaybreaks

\def\km#1{\textcolor{black}{#1}}
\def\jkl#1{\textcolor{black}{#1}}
\def\kmm#1{\textcolor{black}{#1}}

\begin{document}

\title{An Information Theoretic Approach Towards Assessing Perceptual Audio Quality using EEG}
\author{\IEEEauthorblockN{Ketan Mehta\IEEEauthorrefmark{1} and
J\"{o}rg Kliewer\IEEEauthorrefmark{2}, \IEEEmembership{Senior Member, ˜IEEE}}

\IEEEauthorblockA{\IEEEauthorrefmark{1}Klipsch School of Electrical and Computer Engineering, New Mexico State University, Las Cruces, NM 88003}\\
\IEEEauthorblockA{\IEEEauthorrefmark{2}Helen and John C.~Hartmann Department of Electrical and Computer Engineering, New Jersey Institute of Technology, Newark, NJ 07103}
\thanks{This work was supported in part by NSF grant CCF-1065603 and presented in part at the IEEE International Conference on Acoustics, Speech, and Signal Processing (ICASSP), Florence, Italy, May 2014 \cite{Mehta14}.}}
\date{\today}
\maketitle

\begin{abstract}
In this paper, we propose a novel information theoretic model to interpret the entire ``transmission chain'' comprising stimulus generation, brain processing by the human subject, and the electroencephalograph (EEG) response measurements as a nonlinear, time-varying communication channel with memory. We use mutual information (MI) as a measure to assess audio quality perception by directly measuring the brainwave responses of the human subjects using a high resolution EEG. Our focus here is on audio where the quality is impaired by time varying distortions. In particular, we conduct experiments where subjects are presented with audio whose quality varies with time between different possible quality levels. The recorded EEG measurements can be modeled as a multidimensional Gaussian mixture model (GMM). In order to make the computation of the MI feasible, we present a novel low-complexity approximation technique for the differential entropy of the multidimensional GMM. We find the proposed information theoretic approach to be successful in quantifying \km{subjective} audio quality perception, with the results being consistent across different  music sequences and distortion types.
\end{abstract}


\section{Introduction}
\km{An often used metaphor for the human brain is that of an information
  processing device. Our perception of the world around is received as input
  to the brain via sensory nerves. Detecting, interpreting, and responding
  to a stimulus is a multistage process that results in the activation and
  hierarchical interaction of several different regions in the brain. In
  these hierarchies, sensory and motor information is represented and
  manipulated in the form of neural activity, where  neurons transfer
  information about stimulus features by successively transmitting impulse
  trains of electrical or chemical signals. Information in the brain is
  therefore inherently encoded as sequences of neural activity
  patterns. The superposition of these activity patterns can be
  non-invasively  recorded via electroencephalography (EEG) by placing
  sensors on the scalp. In the following we provide a novel information theoretic model to analyze the overall transmission chain comprising of input stimulus generation, processing by the human brain, and the output EEG brain activity, as a time-varying, nonlinear communication channel with memory.}

%

\km{Information theory (IT) provides a stochastic framework which is also
  well suited to characterize and model neural response \cite{spikes,dimitrov2011information,johnson2010information,aghagolzadeh2010synergistic,ostwald2010information,ostwald2012eeg}. In
  fact, soon after Shannon's initial work \cite{shannon2001mathematical},
  the concept of information was applied to calculate the capacity and
  bounds of neural information transfer
  \cite{Stein67,Rapoport,mcculloch1952upper}. 
  In particular, entropy and mutual information (MI)
  \cite{cover2012elements} can be used to quantify how much information a
  neural response contains about a stimulus\cite{Borst,spikes}. In
  \cite{Paninski} the author discusses local expansions for the
  nonparametric estimation of entropy and MI useful for neural data
  analysis.  Further, \cite{panzeri2001unified,panzeri2001role} present a
  power-series-based component expansion for calculating the MI in order to
  analyze information conveyed by the spike timing of neuron firing in the
  somatosensory cortex of a rat. In \cite{dimitrov2003analysis} different
  approximations of Gaussian mixture models are used to bound the MI
  transmission from stimulus to response in the cercal sensory system of a
  cricket. }

\jkl{In contrast to these works which assess information transfer on the neural
level, we focus on the ``high-level'' information flow between stimulus and EEG
sensor observations.} In such a setup MI has also has been successfully
employed in the past for EEG feature extraction and classification purposes
\cite{aviyente2004characterization,kushwaha1992information,wu2007classifying,temko2010speech}. Notably
in \cite{Xu}, MI is used to describe information transmission among
different cortical areas during waking and sleep states using EEG
measurements.  Other researchers have similarly used MI to analyze EEG data
to investigate corticocortical information transmission for pathological
conditions such as Alzheimer's disease \cite{Alzh} and schizophrenia
\cite{Schiz}, for odor stimulation \cite{Odor1,Odor2}, and even for motor
disorders such as Parkinson's disease \cite{wang2009novel}.

For subjective quality testing of audio the current state-of-the-art
approach is Multi Stimulus with Hidden Anchor (MUSHRA) \cite{MUSHRA}. One
characteristic that MUSHRA and most of the other existing audio and video
testing protocols have in common is that each human participant assigns a
single quality-rating score to each test sequence. Such testing suffers from
a subject-based bias towards cultural factors in the local testing
environment and tends to be highly variable. 
In contrast, by using EEG to directly analyze the brainwave patterns we
capture fast, early brain responses that depend only on the perceived
variation in signal quality. Because of these reasons there has recently
been a growing interest in using EEGs to classify human perception of audio
\cite{porbadnigk2010using,creusere2012assessment} and visual quality \cite{scholler2012toward,lindemann2011evaluation,mustafa2012eeg,mustafa2012single}. For example, \cite{creusere2012assessment} investigates the use of a time-space-frequency analysis to identify features in EEG brainwave responses corresponding to time-varying audio quality. Further, \cite{porbadnigk2010using,scholler2012toward} propose to use linear discriminant analysis (LDA) classifiers to extract features based on the P300 ERP component \cite{picton1992p300,sutton1967information} for classifying noise detection in audio signals and to assess changes in perceptual video quality, respectively. Similarly, in \cite{mustafa2012single} the authors employ a wavelet-based approach for an EEG-classification of commonly occurring artifacts in compressed video, using a single-trial EEG. 	




%
\jkl{Our approach here is different compared to above-mentioned studies in
  that we use MI to quantitatively measure how accurately the quality of the
  audio stimulus is transmitted over the brain response channel, where the
  observation at the channel output is given by the EEG measurements. In
  other words, we ask how useful EEG measurements are to assess subjective
  audio quality. How well can the audio quality be blindly estimated (i.e., without
knowing the original audio quality) from the EEG measurements? This could be
useful for example for designing custom hearing aids optimized based on
perceived audio quality or for a subjective assessment of audio streams
which slowly vary their quality with time. For the sake of simplicity and analytical tractability we
  restrict ourselves to the worst case scenario of only two audio quality
  levels (high quality and bad quality).}

\jkl{  We show that the EEG measurements can be modeled as a multidimensional
  Gaussian mixture model (GMM). A direct computation of the MI from the EEG
  measurements is not feasible due to the high dimensionality of the data
  captured by using a large number of EEG sensors. Instead, we present a
  low-complexity approximation technique for the differential entropy of the
  multidimensional GMM based on a Taylor series expansion, which is essential
  for carrying out the  MI computation problem in acceptable time. Additionally, we
  also provide novel analytical differential entropy bounds for the
  one-dimensional GMM case.}  \km{To the best of our knowledge, this is the
  first time that an information theoretic characterization has been used to
  evaluate human perception of audio quality by directly measuring the brain
  activity of the subjects using a high resolution EEG.}

The rest of the paper is organized as follows. \km{In Section II we provide an overview of EEG, the ERP channel, the experiment, and the stimulus audio sequences.} In Section III we analyze the ERP channel using an information theoretic framework and calculate the MI over this end-to-end channel. The results are presented in Section IV, which is followed by a discussion in Section V. We conclude with a summary of our study and future directions in Section VI.


\section{Background}

\subsection{Electroencephalography (EEG)}
\km{EEG is an observation of the electrical activity in the brain recorded over a
  period of time using electrodes placed on the scalp. As mentioned earlier,
  a single EEG electrode measures the sum electric potential resulting from the synchronous activity of several hundred million neurons averaged over tissue masses.
The large spatial scale of EEG recordings make them unsuitable for studying the activity of neurons or small cell assemblies. Instead, EEG provides a robust measure of cortical brain activity and is associated with cognition and behavior.} 


\km{While EEG is affected by stimulus, it does not require one and occurs
  even in the absence of external stimulus, for, e.g., brain activity
  recorded during different sleeping stages. Event-related potentials (ERP)
  typically represent time averaged EEG data recorded specifically in response to a stimulus like light strobes, audio tones, etc. For this reason ERPs are associated with state-dependent brain processing like selective-attention, task context, memory (recognition), and other changes in mental state \cite{nunez2006electric}.}

\km{Typically, EEG studies attempt to tie specific cognitive processes to averaged data. While these extracranial recordings provide estimates of large-scale synaptic activity and are closely related to the endogenous brain state, the most appropriate way in which to combine the data in order to achieve this goal is not clear.
Common methods of analyzing and processing recorded EEG include epoching,
averaging, time-frequency analysis and linear discriminant analysis,
correlation and coherence rely on linear dependencies and assume the EEG
signal to be stationary. In contrast, employing mutual information (MI) is
well suited for modeling nonlinear, non-stationary phenomena like EEGs as it
is based on empirically observed  probability distributions and also performs an implicit averaging.} 


\subsection{ERP Channel}

\begin{figure}[htpb]%
\centering
\subfigure[Communicating over the ERP channel.]{\includegraphics[scale=0.54]{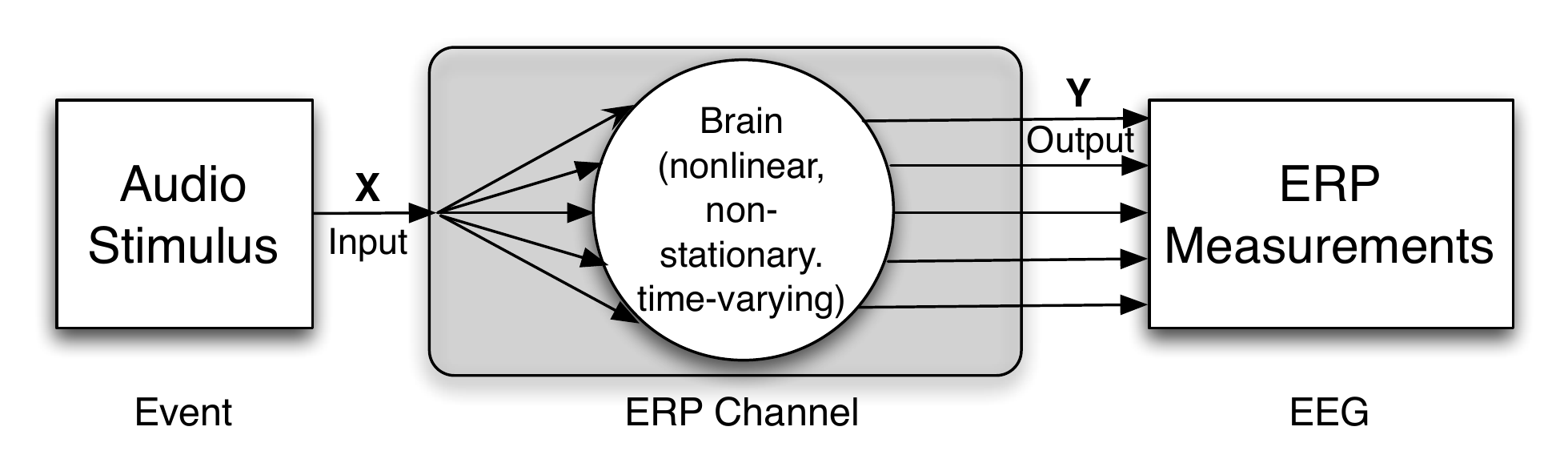}}\\
\subfigure[Single-input multiple-output channel (SIMO) channel.]{\includegraphics[scale=0.38]{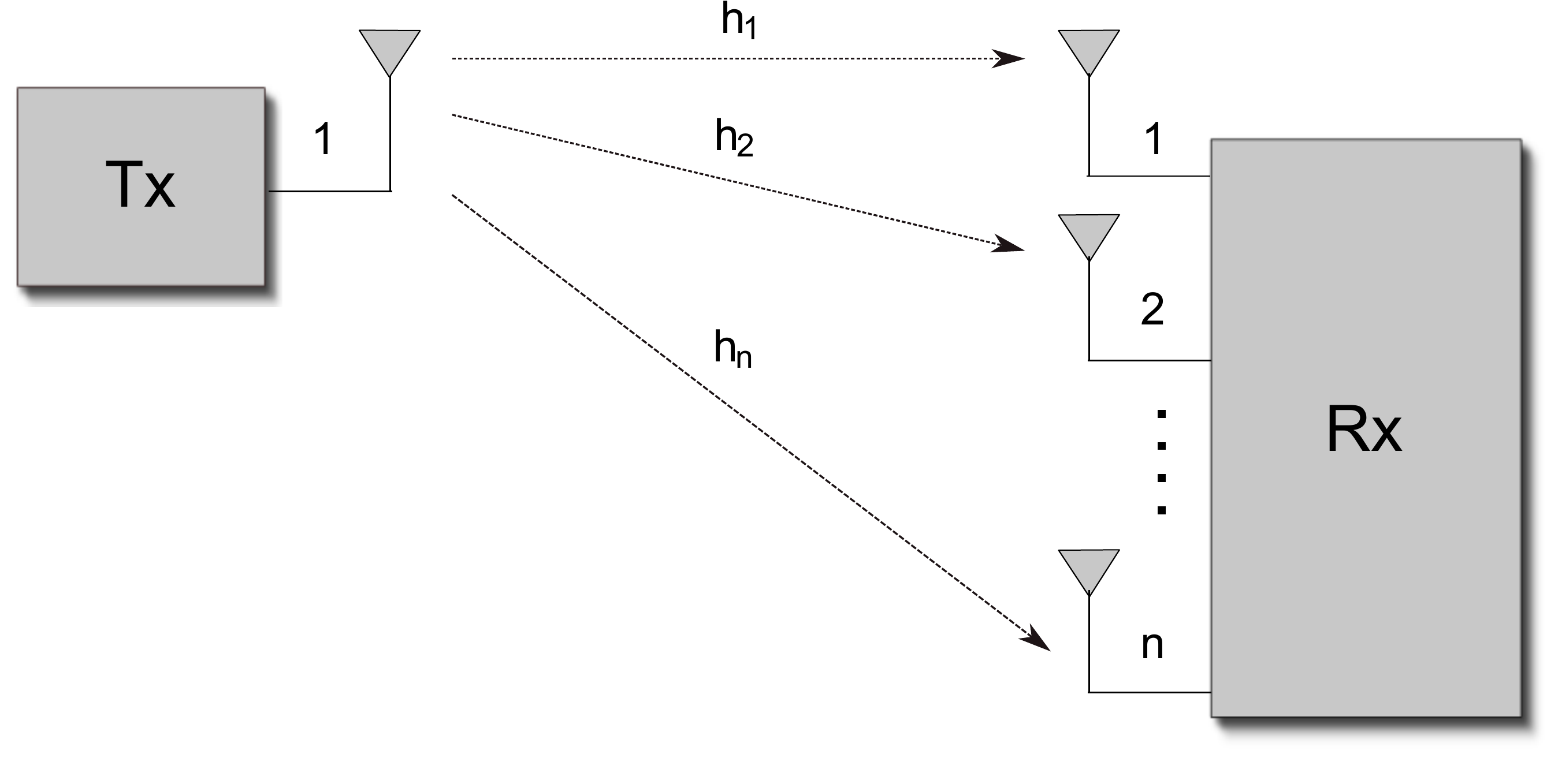}}
\caption{The end-to-end perceptual processing channel is analogous to a SIMO
  communication channel. The input to the channel is the audio quality level and the output is the EEG response.}
\label{fig:ERP}
\end{figure}

\km{Our goal here is to use MI to quantify the information transfer between
  the input audio stimulus and the recorded EEG. In order to achieve this
  goal, we model the overall transmission chain comprising of stimulus generation, processing by the human brain, and the EEG sensors as a communication channel (Fig.~\ref{fig:ERP}(a)). In the neuroscience literature such a channel in general is referred to as the ERP channel \cite{vidal1977real}.}

\km{In our case, the ERP channel is equivalent to a single-input multiple-output (SIMO) channel with unknown characteristics. A SIMO channel model is often used in wireless communications to represent a system with a single transmitting antenna and multiple receiving antennas as shown in Fig.~\ref{fig:ERP}(b). 
In particular, for the ERP channel under consideration, the quality of the audio stimulus represents the (single) input, and the observation at the EEG sensor electrodes on the scalp is the (multiple) output of this channel, respectively.}

\subsection{Experiment}
\label{sec:experiment}
\begin{figure}[htb]
	\centering
		\includegraphics[scale=0.1]{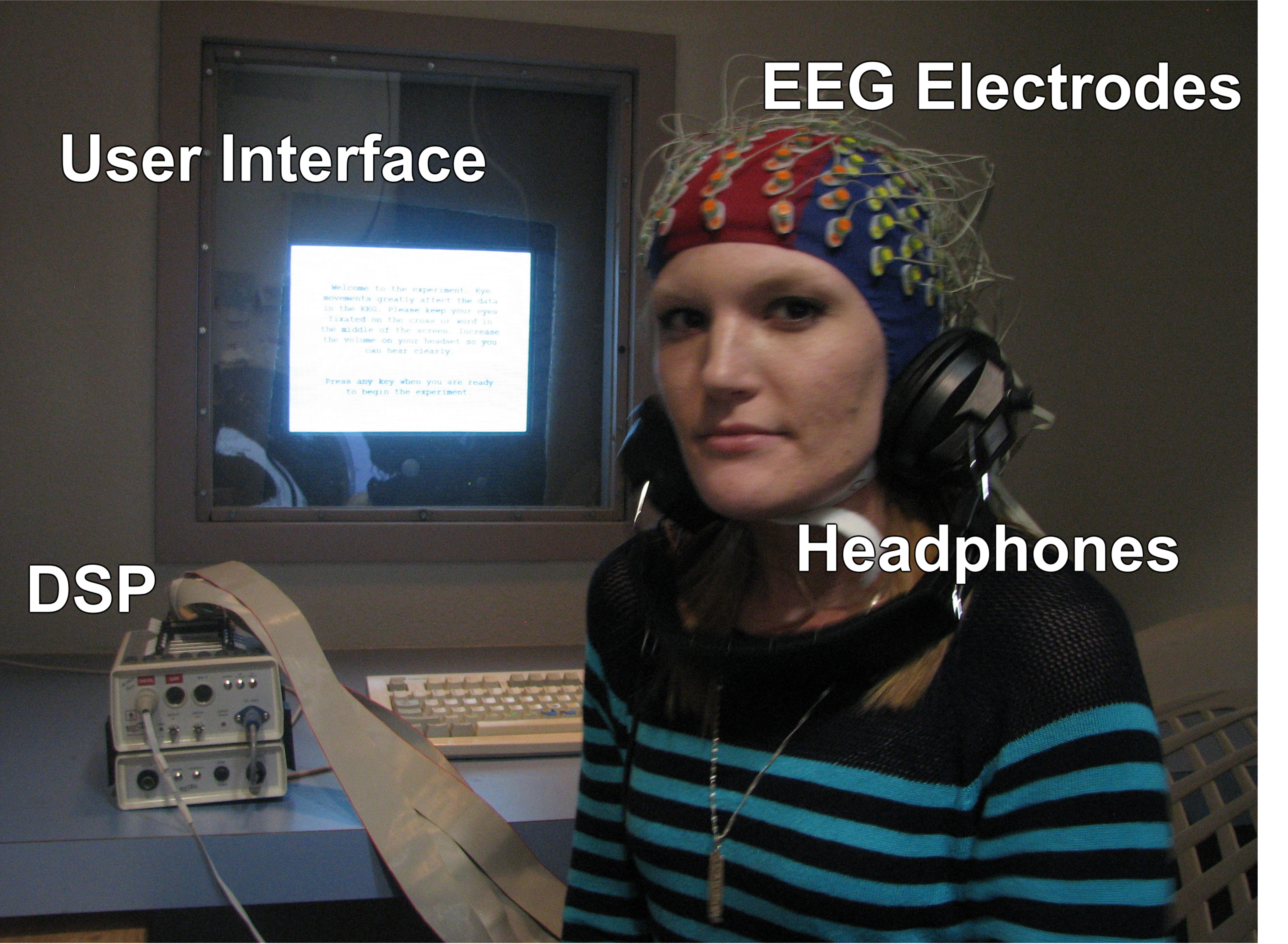}
	\caption{ {The EEG experiment setup.}}
	\label{fig:EEGsetup}
	\vspace{-0.6cm}
\end{figure}

\km{To collect the data required for the study in our experiments, test
subjects were presented with a variety of different audio test sequences whose
qualities varied with time. The EEG activity of the subjects was recorded as
they passively listened to these test sequences via a pair of high fidelity headphones (see Fig.~\ref{fig:EEGsetup}).}

\km{Each trial consisted of a computer controlled presentation of an audio
  test sequence. At the end of the trial, the user interface prompted the
  subject to continue on to the next trial. A 128-channel ActiveTwo Biosemi
  EEG capture system monitored and recorded the brain activity of the
  subjects during the trial. The EEG system comprised of electrodes,
  connecting wires, and a DSP. The DSP re-referenced and amplified the weak
  signals picked up by the electrodes on the scalp and then converted the
  analog signal into a digital signal for storage. The DSP was also connected to the computer and used to synchronize the recordings at the start and end of each trial.}

\km{The trials were conducted in a room specifically built for recording EEGs with a RF shielded testing chamber. A total of 28 test subjects, all with normal hearing capability, participated in the experiment with the majority of them being male.}

\subsection{Stimulus}
\begin{figure}[tb]
	\centering
		\includegraphics[scale=0.85]{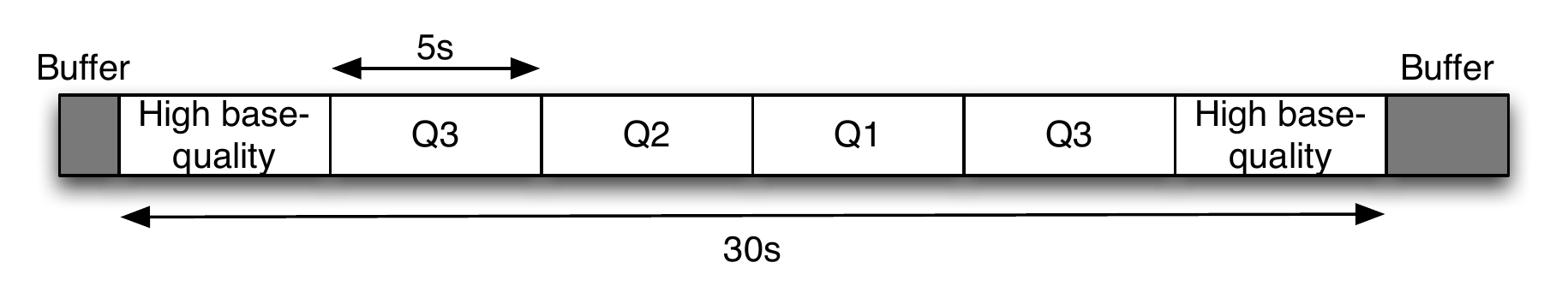}
	\caption{{A 30 second test sequence where the audio quality changes in a time-varying pattern over the whole duration of the sequence. Different possible combinations of quality changes and two distortion types were presented to each subject in a randomized fashion.}}
	\label{fig:aq}
	\vspace{-0.4cm}
\end{figure}

\jkl{All the audio test sequences used as stimulus were created from three
  fundamentally different base-sequences. The base-sequences were sampled at
  44.1\,kHz with a precision of 16 bits per sample. Two different types of
  distortions were implemented -- scalar quantization and frequency band
  truncation -- with each distortion type in itself having three quality
  levels of impairment ``Q1'', ``Q2'', and ``Q3'', respectively (see
  Table \ref{tab:audioquality}). Thirty second long test sequences were created from the
  base-sequence by selecting one of the two distortion types and by applying
  the three possible impairment levels in a time-varying pattern over the
  whole duration of the sequence. Fig.~\ref{fig:aq} shows one of the many
  possible test sequences used in the study. The criterion and use of these
  test sequences and distortion quality levels as a metric for quality
  perception is described in detail in \cite{creusere2011assessing}.}

\kmm{Note that despite the subjects were
  presented with all different quality levels in our listening tests, in our analysis we focus only on the high base-quality and the ``Q3'' degraded quality audio as inputs to the ERP
    channel. This addresses the worst-case quality change and keeps the
    problem analytically and numerically tractable. This choice is also
    supported from a subjective quality assessment viewpoint as it easier to
    distinguish between `good' and `bad' audio, as opposed to `medium-good'
    or `medium-bad' qualities. }

\km{Typically, responses to a stimulus observed by EEG measurements are
  averaged over different trials to remove inconsistencies between these
  trials \cite{freeman2012imaging}.} Evaluating the response to repeated
trials provides a measure of discriminability on the given task, allowing to
reliably judge which state changes. Multiple presentations of both the
impaired test sequences and the unimpaired reference base sequences were made in a randomized fashion. Also, to remove any possible bias, all possible combinations of quality changes, distortion types, and base sequences were presented to each subject.

\renewcommand{\arraystretch}{1.2}
\begin{table}[tbp]
    \centering
	\small
	\caption{Different quality levels presented to the subject during
          the course of an audio trial. To generate the distortion, each of
          these base-sequences were passed through a 2048-point modified discrete cosine transform (MDCT) and either the frequency truncation or the scalar quantization was applied to the coefficients prior to reconstruction \cite{creusere2011assessing}.}
	\begin{tabular}{@{}r|c|c@{}}
			\toprule
			Quality Level & Freq. Truncation & Scalar Quantization \\
					& Low Pass Filter & No.~of Significant Bits Retained \\
			\midrule
			Q1 	& 	4.4 KHz 	&	4	 \\
			Q2 	& 	2.2 KHz 	&	3	 \\
			Q3 	& 	1.1 KHz		&	2	 \\ 
			\bottomrule
	\end{tabular}
	\label{tab:audioquality}
\end{table}

\subsection{EEG Data} \label{sec:ROI}
The EEG system captured data on each of the 128 spatial channels, sampled at
1024\,Hz. An average re-referencing and baseline removal was performed to
remove any DC noise component, which is often introduced over time into the
EEG data usually as a result of steady brain activity, but also due to muscle tension, signal fluctuation, or other noise
sources. Finally, the data was passed through a high-pass filter with a
cut-off frequency at 1\,Hz, and a transition bandwidth of 0.2\,Hz.

\begin{figure}[tb]
	\centering
		\includegraphics[scale=0.27]{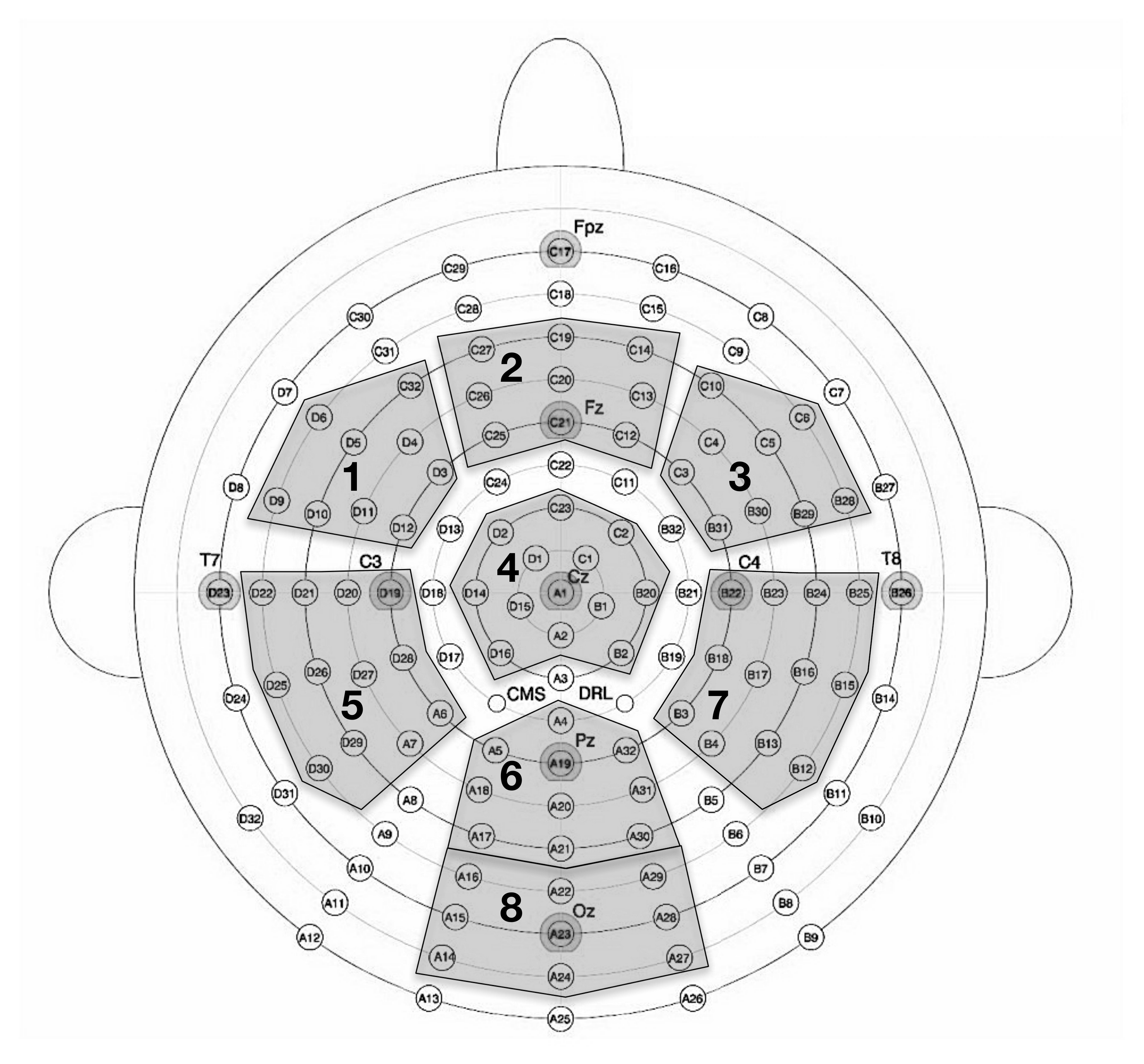}
	\caption{{Schematic representation of the 128-channel EEG system. The electrodes are grouped into eight regions of interest (ROI) to effectively map the activity across different regions of the brain.}}
	\label{fig:biosemi}
	\vspace{-0.4cm}
\end{figure}

We notice that for each trial we have a very large multidimensional
data-set from which to attempt and extract signals corresponding to changes
in the human perception of audio quality. 
To better manage the large amount of collected data while effectively mapping the activity across different regions of the brain, we suggest grouping the 128 electrodes of the
EEG-system into specific regions of interest (ROI). 
The grouping scheme we use is shown in Fig.~\ref{fig:biosemi}, and is similar to the one used in \cite{gruber2006brain}. While a large number of potential grouping schemes are possible, this scheme is favored for our purposes as it efficiently covers all the cortical regions
(lobes) of the brain with a relatively low number of ROIs. For example, regions 5 and 6 cover the temporal lobes, while region 7 spans the parietal lobe. Also, the nearly symmetrical distribution of the ROIs in this scheme allows for an easy
mapping and comparison of EEG activity between different cortical areas.

\section{Information Theoretic Analysis} \label{sec:3}

\textit{Notation:} Throughout this paper, random variables are denoted by uppercase letters $X$ and their (deterministic) outcomes by the same letter in lowercase $x$. Probability distributions are denoted by $p(x)$. A continuous random process is denoted using a time index $Y(t)$. Multivariate random variables are denoted with an underlined uppercase letters $\underline{Y}$. Deterministic vectors are represented as smallcase underlined letters $\underline{\mu}\,$, while matrices are denoted with boldface $\textbf{C}$. The expected value of a function is denoted as $\mathbb{E}[\cdot]$.

\subsection{ERP Channel}\label{sec:31}
\kmm{In our discussion so far, we have shown that the ERP channel under
  consideration (see Fig.~\ref{fig:ERP}) is analogous to a SIMO communication channel.}
The capacity of SIMO channels has been previously analyzed in
\cite{telatar1999capacity}. However, since determining the capacity
achieving input distribution for the ERP channel is virtually impossible to
obtain in an experimental setup, we focus on analyzing the MI for a
uniform input distribution (symmetric capacity) as outlined below. Further,
since the ERP channel is non-linear, non-stationary, potentially time
varying, and does not have an explicitly defined channel characteristic, the
expressions stated in \cite{telatar1999capacity} for the direct computation
of its MI are not applicable.  We begin our analysis therefore by taking a
closer look at the only known parameters of our model, \textit{viz.} the
input and the output random variables.

The input random variable $X$ of the ERP channel is uniformly distributed
over a set of class indices $\mathcal{X}$ which describe the quality of the
stimulus sequences at any given time interval. As discussed in
Section~\ref{sec:experiment} we restrict ourselves to $|\mathcal{X}|=2$. Then, the audio quality of the input sequence can  be represented as an equiprobable Bernoulli distribution
\begin{align}
X = \begin{cases} x_1, & \,\mbox{if the input stimulus is of high quality,}\\
			    x_2, & \,\mbox{if the input stimulus is of degraded quality level `Q3',}\end{cases} \notag
\end{align}
with probabilities
\begin{align}
	p(x_1)=p(x_2)=1/2.
\end{align}

The output of the channel is given by the set $\mathcal{Y}^n\subset\mathbb{R}^n$ containing all possible values of the EEG potential at any given time interval. For a total of $n$ ROIs we therefore get a (multivariate) output vector of random processes $\underline{Y}(t) = (Y_1(t), \dots ,Y_n(t))$. To reduce the variance of the ERP channel measurements we consider every electrode in the
$i$-th ROI as an independent realization of the same random process $Y_i(t)$, $i=1,\dots,n$. This is supported by our grouping scheme wherein all electrodes of a ROI are located within close proximity of one another and capture data over the same cortical region. Further, for tractability we assume the random process to be stationary within the five second non-overlapping blocks of a trial with constant audio-quality. The probability function $p_{\underline{Y}}(\cdot)$ is therefore not dependent on time and can be written as $p_{\underline{Y}}\left(\underline{y}(t)\right)=p_{\underline{Y}}(\underline{y})$ without any loss of generality. Note that this assumption does not rule out any non-stationary behavior between sections of different audio quality within the same trial.  

\begin{figure}[tb]
	\centering
		\includegraphics[scale=0.48]{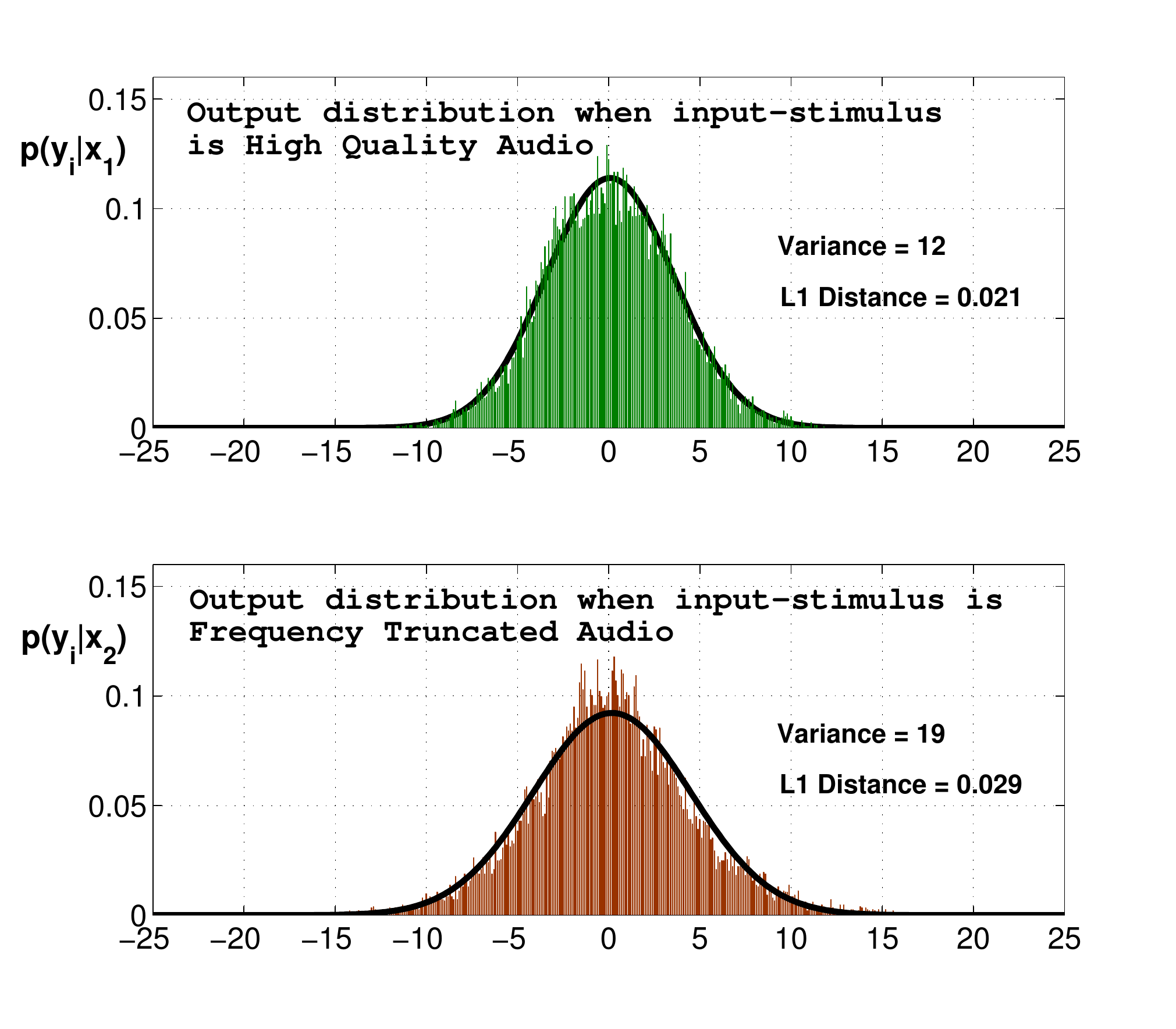}
	\caption{{Marginal conditional distributions $p(y_i|x_1)$ and $p(y_i|x_2)$ of subject S4 over a single ROI, for high quality and frequency distorted audio input-stimulus, respectively. The Gaussian fit is obtained by using an estimator that minimizes the $L_1$ distance between the fitted Gaussian distribution and the histogram data.}}
	\label{fig:hist}
	\vspace{-0.3cm}
\end{figure}

The relationship between the input and output of the ERP channel is characterized by its conditional probability distribution. Since the input can take on two distinct values, there are two marginal conditional distributions $p(y_i|x_1)$ and $p(y_i|x_2)$ corresponding to any given ROI. Fig.~\ref{fig:hist} shows the marginal conditional distributions obtained via
histogram measurements of a single ROI output over time. A detailed
inspection using different subjects, input sequences, and ROIs allows us to
assert two important facts about this distribution. \textit{First}, the conditional distribution converges to a Gaussian with zero mean. The potential recorded at the EEG electrode at any given time-instant can be considered as the superposition of responses of a large number of neurons. Thus, the distribution of a sufficiently high number of these trials taken at different time instances converges to a Gaussian distribution as a result of the Central Limit Theorem (CLT). It then follows directly from the CLT that the probability distribution for $n$ ROIs will also converge to a $n$-dimensional multivariate Gaussian distribution.
\textit{Second}, we observe from Fig.~\ref{fig:hist} that there is a difference between the variances of the distributions $p(y_i|x_1)$ and
$p(y_i|x_2)$. This indicates that $Y_i$ is dependent on $X$, i.e., the EEG
data is related to and contains ``information" about the input stimulus, which is characterized by the MI $I(X;Y_i)$ as a measure of the information transfer over the ERP channel. 

\subsection{Entropy and Mutual Information}\label{sec:32}
The entropy\footnote{In the following, we simply use the denomination ``entropy" even when referring to differential entropy for the case of continuous random variables.} of a (continuous or discrete) random variable $X$ with a probability distribution $p(x)$ quantifies the amount of information in $X$, and is defined as \cite{cover2012elements}
\begin{equation}
	h(X) = \mathbb{E}[-\log p(x)]. 
\end{equation}
Unless otherwise specified we will take all logarithms to base 2, thus measuring the entropy and MI in bits. For two random variables $X$ and ${Y}$ with a joint probability density function $p(x,y)$ the MI between them is defined as \cite{cover2012elements}
\begin{equation}
I(X;Y) = \mathbb{E}_{XY}\left[ \log\frac{p(x,y)}{p(x)p(y)}\right].
\end{equation}
In this work we are interested in the MI between the ERP channel input $X$ and output $\underline{Y}$. Therefore, a moderate-to-high value for $I(X;\underline{Y})$ indicates that the EEG data output is related to the input stimulus, i.e., contains information about the audio quality. 

\begin{figure}[htpb]%
\centering
\subfigure[]{\includegraphics[scale=0.42]{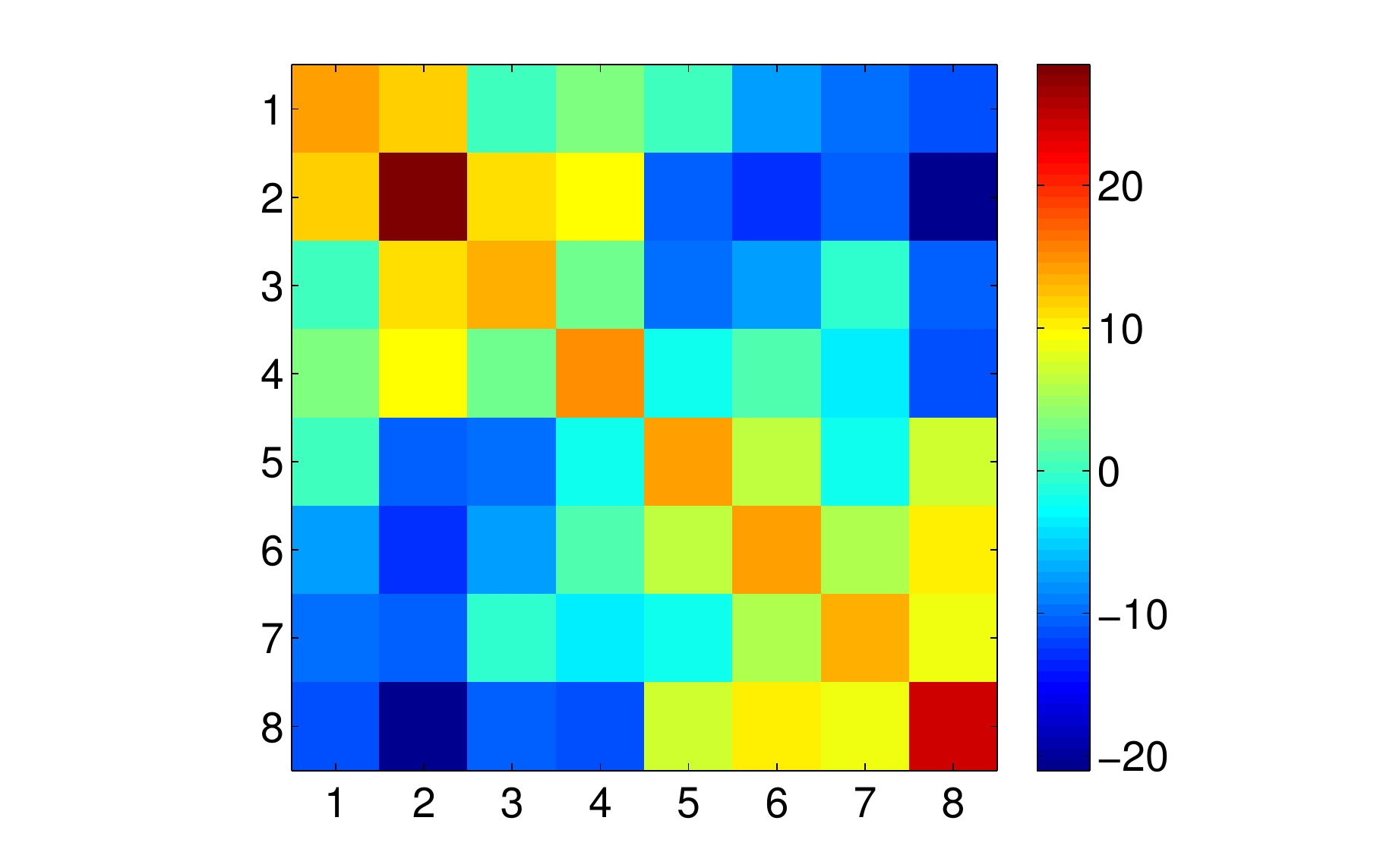}}
\subfigure[]{\includegraphics[scale=0.42]{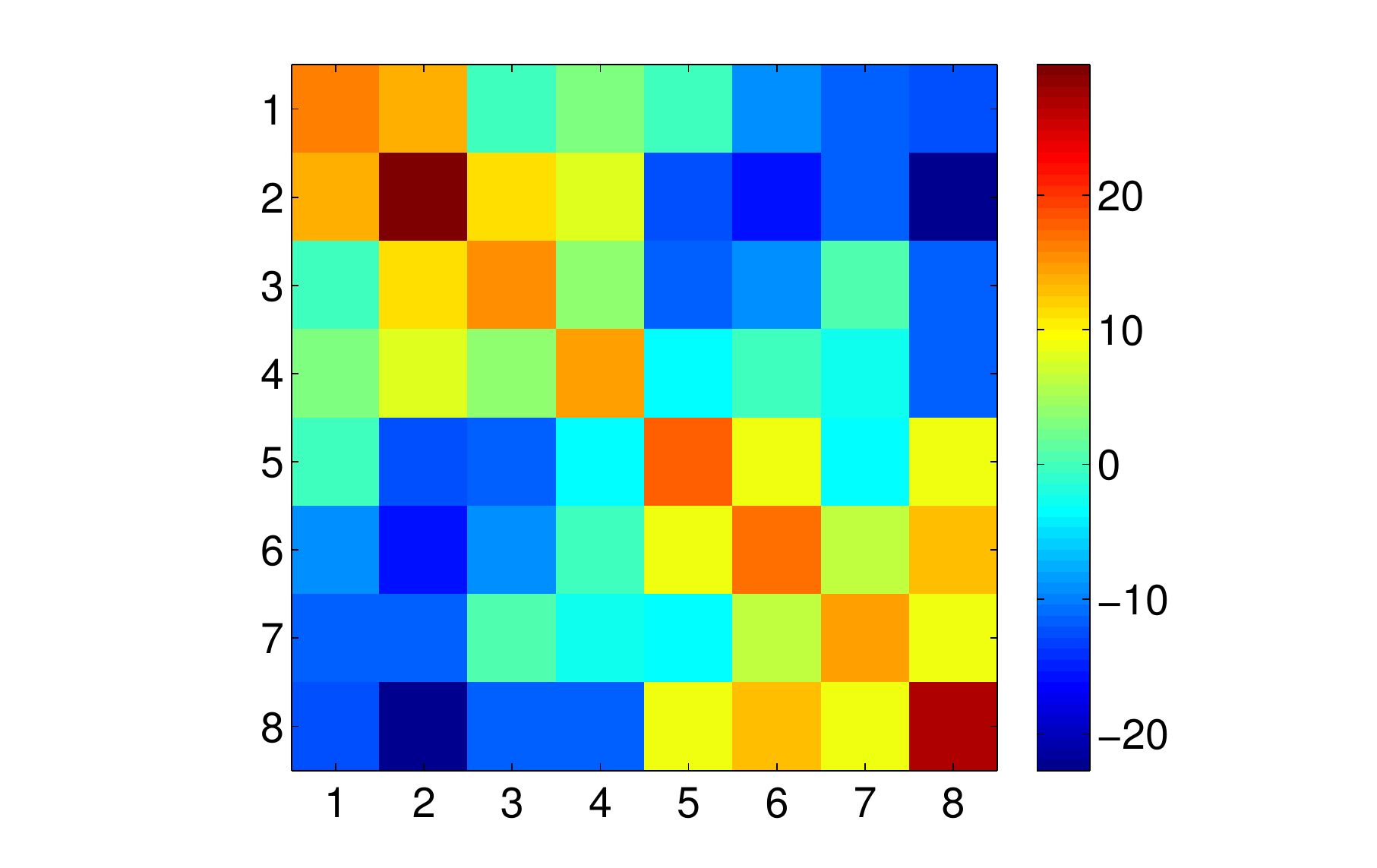}}
\caption{The covariance matrices corresponding to the mulitvariate Gaussian conditional distributions (a). $p(\underline{y}|x_1)$ and (b). $p(\underline{y}|x_2)$ of subject S4 over $n$=8 ROIs, indicating the high correlation between the different regions.}
\label{fig:covmatrix}
\end{figure}

If we consider the conditional probability distributions corresponding to $n$ ROIs chosen simultaneously, then each of the distributions $p(\underline{y}|x=0)$ and $p(\underline{y}|x=1)$ are $n$-dimensional Gaussian distributions, with $\mathbf{C_1}$ and $\mathbf{C_2}$ being the $n\times n$ covariance matrices of each of the distributions respectively. Fig.~\ref{fig:covmatrix} shows the covariance matrices of a subject over $n=8$ regions for a single trial.
The conditional entropy $h(\underline{Y}|X)$ is then given by
\begin{eqnarray}
	h(\underline{Y}|X) 	&\triangleq& -\sum\limits_{x\in X} p(x)\!\!\int\limits_{-\infty}^\infty p(\underline{y}|x)\log{p(\underline{y}|x)\,d\textbf{y}} \nonumber \\
					&=& \frac{1}{4} \left\{	\log(2\pi e)^{n}|\mathbf{C_1}|	+	\log(2\pi e)^{n}|\mathbf{C_2}|	\right\},
\end{eqnarray}
where we have used the closed-form expressions for the differential entropy of a multivariate Gaussian distribution \cite[Theorem 8.4.1]{cover2012elements}. 
Using the law of total probability we can write $p(y)$ in terms of the conditional probabilities as
\begin{align} \label{eq:4}
	p(\underline{y}) 	&=  \sum\limits_{x\in X} p(\underline{Y}\!\!=\! \underline{y}|x)p(x) \notag \\
				&=  \frac{1}{2} \left\{	p(\underline{y}|x_1)	+	p(\underline{y}|x_2)	\right\}.
\end{align}
\jkl{As both $p(\underline{y}|x_1)$ and $p(\underline{y}|x_2)$ are
Gaussian distributed, the resulting random vector $\underline{y}$ is a
multivariate Gaussian mixture with a distribution $p(\underline{y})$ given
in \eqref{eq:4}.} 
Therefore, the entropy of $\underline{Y}$ is given by
\begin{align} \label{eq:gaussmix}
\!h(\underline{Y})\!=\! -\frac{1}{2}\!\!\int\limits_{\mathbb{R}^n}\!\!\left[p(\underline{y}|x_1)	\!+\!	p(\underline{y}|x_2)\right]\!\cdot\!\log\left(\frac{1}{2}\left[p(\underline{y}|x_1)	\!+\!	p(\underline{y}|x_2)\right]\right)d\underline{y}.
\end{align}
The MI between the output EEG data and the input audio stimulus can then be calculated using
\begin{align}
I(X;\underline{Y}) &= h(\underline{Y}) - h(\underline{Y}|X).
\end{align}
\subsection{Entropy approximation} \label{sec:33}
The entropy of the output random variable (\ref{eq:gaussmix}) is the entropy of a mixture of two Gaussian distributions. It turns out, to the best of our knowledge, that there is no closed form solution for the entropy of a Gaussian mixture model (GMM) which is also a recurring open problem in the literature. The expression for the entropy of a GMM consists of a multiple integral over a logarithm of a sum of exponential functions which makes it difficult to formulate a general closed-form analytical solution. In the absence of an analytical solution it is usually common to use a numerical method to calculate a sufficiently accurate estimate of the solution. However, as the dimensionality of the Gaussian multivariate random variable increases, it becomes computationally infeasible to evaluate the $n$-order multiple integral using numerical integration. This drawback becomes even more apparent in our case where several test subjects each with multiple test sequences must be processed.
In keeping with our primary goal of evaluating the overall information transfer over the ERP channel, we take a closer look at this open problem in the following subsections and propose two novel low-complexity methods to estimate the differential entropy of a GMM.

\subsubsection{Single region entropy bounds}\label{sec:331}
In the following we derive new, tight upper and lower bounds for the entropy of a single ROI. The critical factor in the entropy calculation of the Gaussian mixture is the logarithm of the sum of individual distributions. By employing the Jacobian logarithm we obtain an alternate expression for the logarithm of a sum
\begin{align} \label{eq:jacob1}
\ln(e^a+e^b)=\max{(a,b)} + \ln{(1+e^{-|a-b|})}\, ,
\end{align}
where $a$ and $b$ are constants. The expression in (\ref{eq:jacob1}) can be bounded as
\begin{align} \label{eq:jacob2}
\max{(a,b)} < \ln(e^a+e^b) \leq \max{(a,b)} + \ln{2}\, ,
\end{align}
where equality on the upper bound is achieved when $a=b$. Substituting
\begin{align}
					e^a 	 &\rightarrow \frac{1}{\sigma_1}\exp{\frac{-y^2}{2\sigma_1^2}}  \notag  \\
	\mbox{and}\quad   	e^ b  &\rightarrow \frac{1}{\sigma_2}\exp{\frac{-y^2}{2\sigma_2^2}} \, ,
\end{align}
we obtain the following bounds
\begin{multline} \label{eq:jacob3}
	 \max\left(-\frac{y^2}{2\sigma_1^2}+\ln\sigma_1\enspace,\enspace-\frac{y^2}{2\sigma_2^2}+\ln\sigma_2\right)
	\;\;\;\leq\;\;\;
	\ln \left(	\frac{e^{\frac{-y^2}{2\sigma_1^2}}}{\sigma_1} + 	\frac{e^{\frac{-y^2}{2\sigma_2^2}}}{\sigma_2}\right) \\
	\leq\;\;\;
	\max\left(-\frac{y^2}{2\sigma_1^2}+\ln\sigma_1\enspace,\enspace-\frac{y^2}{2\sigma_2^2}+\ln\sigma_2\right) + \ln{\left(1 + \frac{\sigma_2}{\sigma_1}\right)}\, .
 \end{multline}
Here $\sigma_1^2$, $\sigma_2^2$ are the variances of the two Gaussian distributions $p(y|x_1)$ and $p(y|x_2)$, respectively, chosen such that  $\sigma_2 > \sigma_1$.
We can now use the lower bound in \eqref{eq:jacob3} to calculate the upper bound for the maximum entropy $h(Y)$ for a single ROI,
\begin{multline} \label{eq:jacobbound}
	h(Y) 	\leq
				-\frac{1}{2}\int\limits_{-\infty}^\infty [p(y|x_1)	+	p(y|x_2)] \cdot {\max\left( \frac{y^2}{2\sigma_1^2}+\ln{2\sqrt{2\pi}\sigma_1} \enspace , \enspace \frac{y^2}{2\sigma_2^2}+\ln{2\sqrt{2\pi}\sigma_2} \right) }\, dy \\
				= \int\limits_{0}^{\lambda} [p(y|x_1)	+	p(y|x_2)]\cdot\left(\frac{y^2}{2\sigma_1^2}+\ln{2\sqrt{2\pi}\sigma_1}\right)\, dy \\
					+\int\limits_{\lambda}^{\infty} [p(y|x_1)	+	p(y|x_2)]\cdot \left(\frac{y^2}{2\sigma_2^2}+\ln{2\sqrt{2\pi}\sigma_2}\right)\, dy,
\end{multline}
where $\pm\lambda$ are the points of intersection of the two component Gaussian densities of the GMM. The points of intersection can be found by equating the two Gaussian densities and solving,
\begin{align} \label{eq:jacob4}
			&(1/\sqrt{2\pi}\sigma_1)e^{\frac{-\lambda^2}{2\sigma_1^2}} = (1/\sqrt{2\pi}\sigma_2)e^{\frac{-\lambda^2}{2\sigma_2^2}}  \notag \\
	\implies 	& \lambda = \frac{\sigma_1^2\sigma_2^2\cdot\left(\ln{\sigma_1^2}-\ln{\sigma_2^2}\right)}{\sigma_1^2 - \sigma_2^2}.
\end{align} 
The two integrals in (\ref{eq:jacobbound}) have a closed form solution and can be solved using the expressions for truncated normal distributions as
\begin{multline} \label{eq:jacob5}
	h(Y) 	\leq  \frac{1}{2}\left( -\frac{\lambda}{\sigma_1}\,Q\left(\frac{\lambda}{\sigma_1}\right) + \frac{1}{c_1} \right)	+ \ln({2\sqrt{2\pi}\sigma_1})\cdot\left[\frac{1}{c_1}+\frac{1}{c_2}\right]	+ \frac{\sigma_2^2}{2\sigma_1^2}\left( -\frac{\lambda}{\sigma_2}\,Q\left(\frac{\lambda}{\sigma_2}\right) + \frac{1}{c_2} \right) \\
	+ \frac{\sigma_1^2}{2\sigma_2^2}\left( -\frac{\lambda}{\sigma_1}\,Q\left(\frac{\lambda}{\sigma_1}\right) + \frac{1}{d_1} \right)	+ \ln({2\sqrt{2\pi}\sigma_1})\cdot\left[\frac{1}{d_1}+\frac{1}{d_2}\right]	+ \frac{1}{2}\left( -\frac{\lambda}{\sigma_2}\,Q\left(\frac{\lambda}{\sigma_2}\right) + \frac{1}{d_2} \right),		
\end{multline}
where
\begin{align} \label{eq:jacob6}
			& c_1 = \frac{1}{\Phi(\frac{\lambda}{\sigma_1})-\Phi(0)}\,,\quad c_2 = \frac{1}{\Phi(\frac{\lambda}{\sigma_2})-\Phi(0)}\,, \notag \\
			& d_1 = \frac{1}{1-\Phi(\frac{\lambda}{\sigma_1})}\,,\quad\quad d_2 = \frac{1}{1-\Phi(\frac{\lambda}{\sigma_2})}\,,
\end{align} 
and $\Phi(\cdot)$ is the standard normal cumulative density function, and $Q(\cdot)$ is the standard normal probability density function. A similar approach can be used to calculate a lower bound on the minimum entropy of $h(Y)$ by using the upper bound in \eqref{eq:jacob3}.

This method is successful in providing tight bounds for the the entropy of a mixture of two scalar normal random variables, i.e., the entropy for an individual ROI. A limitation of this method however is that it is not effective for evaluating multivariate normal distributions and is therefore unsuitable for evaluating the entropy of multiple ROIs chosen simultaneously. In particular, a mixture of a $n$-dimensional normal distribution will have intersections in a $n$-dimensional space and the integral in (\ref{eq:jacobbound}), if it even exists, will be extremely complex.

\subsubsection{Multiple region entropy approximation}\label{sec:332}
In order to address these shortcomings for higher dimensional EEG outputs $\underline{Y}$ we propose an approximation for the entropy by performing a component-wise Taylor series expansion of the logarithmic function
\cite{huber2008entropy}. This approximation makes no prior assumptions and
is suitable in general for estimating the entropy of any given
multidimensional GMM.

Let $p_i(\underline{z})\sim \mathcal{N}(\underline{z};\underline{\mu}_i,\mathbf{C}_i)$ be the probability distribution for the $i$-th component of the GMM associated with the $n$-dimensional Gaussian distribution
$\mathcal{N}(\underline{z};\underline{\mu}_i,\mathbf{C}_i)$ with mean
$\underline{\mu}_i\in \mathbb{R}^n$ and covariance matrix
$\mathbf{C}_i\in\mathbb{R}^{n\times n}$. The probability distribution of the GMM is then given by 
\begin{align} \label{taylor1}
	p(\underline{z}) &= \sum\limits_{i=1}^{L} w_i p_i(\underline{z}),
\end{align}
where $L$ is the number of mixture components, and $w_i$ denotes the weight of the $i$-th component of the GMM, respectively. Also, let
\begin{align} \label{taylor2}
	f(\underline{z}) 	&= \log p(\underline{z}) = \log\left(\sum\limits_{j=1}^{L} w_j p_j(\underline{z})\right).				
\end{align}
If we then use the Taylor series to expand the function $f(\underline{z})$ around $\underline{\mu}_i$,  we obtain
\begin{align} \label{taylor3}
	f(\underline{z}) 	=  f(\underline{\mu}_i) + \frac{f^{'}(\underline{\mu}_i)}{1!}(\underline{z}-\underline{\mu}_i)
												+ \frac{f^{''}(\underline{\mu}_i)}{2!}{(\underline{z}-\underline{\mu}_i)}^2 + \ldots
\end{align}\\*
The odd central moments of a Gaussian distribution are zero \cite{shynk2012probability,oddorder1}. \km{Therefore, all the odd order terms in the Taylor series expansion of the entropy of the Gaussian mixture are also zero, resulting in the following expansion}
\begin{align}  \label{taylor4}
	h(\underline{z}) 	&= -\!\int\limits_{\mathbb{R}^n}\!\!p(\underline{z})\log p(\underline{z}) \,d\underline{z} = -\!\int\limits_{\mathbb{R}^n}\!\!p(\underline{z})f(\underline{z}) \,d\underline{z} \notag \\
	 						&=  -\sum\limits_{i=1}^{L}\int\limits_{\mathbb{R}^n}\!\!w_i p_i(\underline{z})\cdot \left\{f(\underline{\mu}_i) + \frac{f^{''}(\underline{\mu}_i)}{2!}{(\underline{z}-\underline{\mu}_i)}^2 + \ldots\right\} d\underline{z}.	
\end{align}
The Gaussian mixture is a smooth function and the derivatives of the function $f(\underline{z})$ will therefore always exist. The Taylor series approximation is exact only if an infinite order of terms are considered, and the accuracy of the approximation therefore depends on the number of expansion terms used. We provide closed-form expressions for the zeroth-order, second-order, and fourth order terms of the Taylor series expansion in the Appendix at the end of this paper. 

The Taylor series in \eqref{taylor3} is expanded around the mean of the distribution. As the variance of the distribution increases, the data set is spread further away from the mean and as a result a larger number of higher-order terms are required to reduce the approximation error. Also, if one of the distributions has a relatively high variance compared to the other distributions in the mixture, then the corresponding $n$-dimensional pdf is skewed heavily along that particular axis.

\begin{figure}[htbp] 
\begin{center}
\centerline{\includegraphics[scale=0.28]{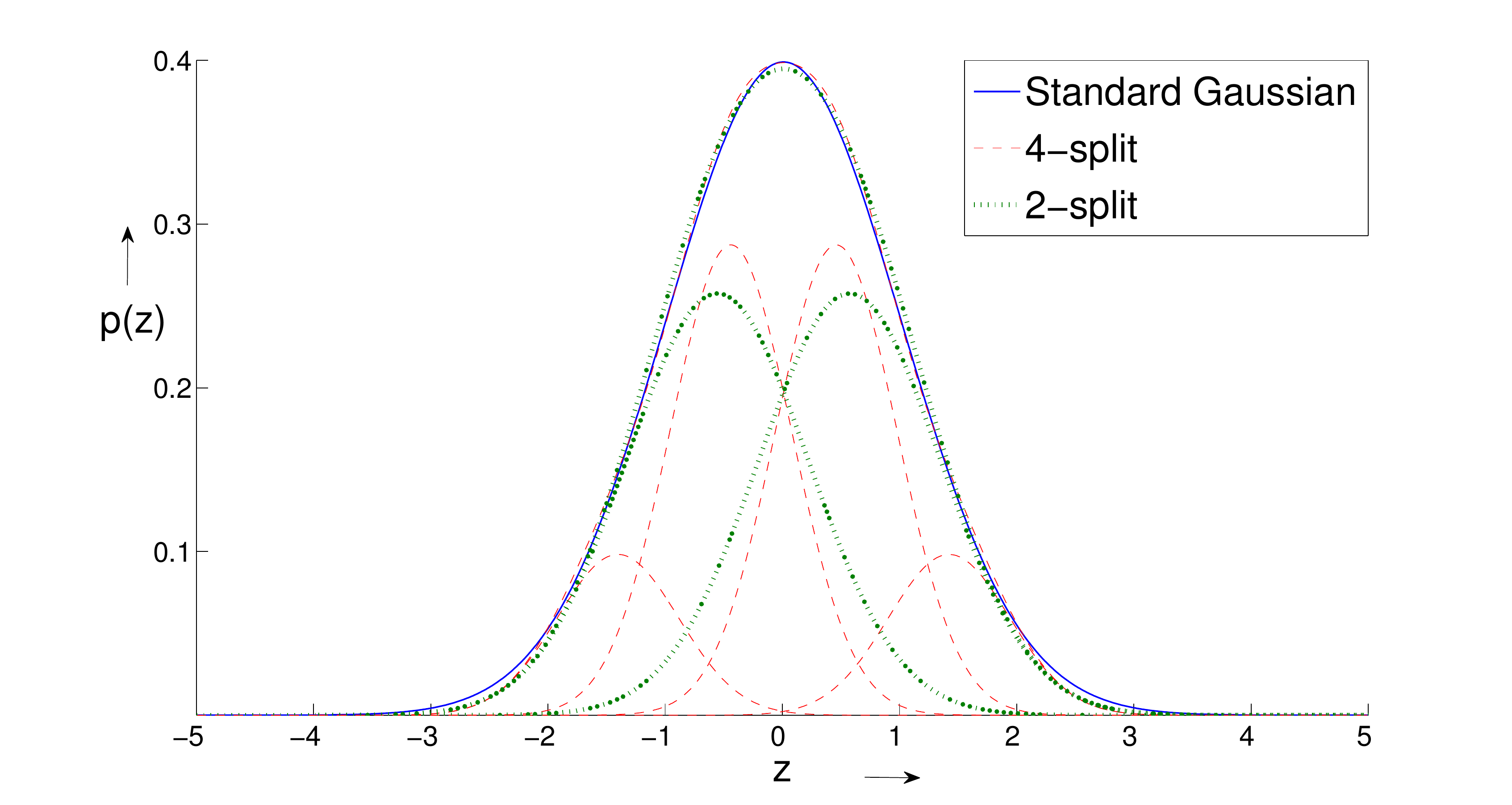}}
\caption{Variance split of a standard Gaussian using the splitting library in Table \ref{table:split}.}
\label{fig:split}
\end{center}
\vspace{-0.5cm}
\end{figure}

Calculating a large number of higher order terms is complex and
computationally demanding. In order to obtain a high accuracy approximation
while keeping the number of expansion terms constant, we propose using a
\textit{variance splitting} approach
\cite{huber2008entropy,hanebeck2003progressive}. In this approach we split
and replace the high variance Gaussian component with a mixture of
Gaussians, each with a substantially lower variance than the original. The
$i$-th component with $i=1,2,\dots,L$ for splitting is identified and the corresponding covariance matrix $\mathbf{C}_i$ is diagonalized as
\begin{align}
	\mathbf{D}_i = \mathbf{\Sigma}_i^T \mathbf{C}_i \mathbf{\Sigma}_i\,\,,
\end{align}
such that
\begin{align}
\mathbf{D}_i = \mbox{diag}\left( \lambda^{(i)}_1, \lambda^{(i)}_2 , \ldots ,
  \lambda^{(i)}_n	\right) 
\end{align}
is the diagonal matrix containing the eigenvalues of the matrix $\mathbf{C}_i$, and $\mathbf{\Sigma}_i$ is the diagonalization matrix whose columns are the eigenvectors of $\mathbf{C}_i$. The $i$-th mixture component is then split along the principal axis of its longest ellipsoid into $K$ sub-components as
\begin{align}
	w_i\cdot p_i(\underline{z}) = \sum\limits_{k=1}^{K}w^{(i)}_k\cdot w_i\cdot p^{(i)}_k(\underline{z}) ,
\end{align}
where   $\bar{w}_k,\bar{\mu}_k$, and $\bar{\sigma}_k$ are 
splitting parameters. Further, we define 
\begin{align}
&w^{(i)}_k \triangleq  \bar{w}_k w_i \quad \mbox{with} \quad \sum\limits_{k=1}^{K} w^{(i)}_k = w_i, \notag \\
&\underline{\mu}^{(i)}_k \triangleq  \underline{\mu}_i + \sqrt{\lambda^{(i)}_d}\cdot\bar{\mu}_k\cdot[0,\ldots,1,\ldots,0]^T, \notag \\
&\mathbf{D}^{(i)}_k \triangleq  \mbox{diag}\left( \lambda^{(i)}_1,\ldots , \lambda^{(i)}_{d-1},\bar{\sigma}_k^2,\lambda^{(i)}_{d+1},\ldots, \lambda^{(i)}_n	\right),
\end{align}
where $\lambda^{(i)}_d$ is the largest eigenvalue of $\mathbf{C}_i$. The new covariance matrices, $\mathbf{C}^{(i)}_k$, for the sub-components  are obtained by rotating back
\begin{align}
	\mathbf{C}^{(i)}_k = \mathbf{\Sigma}_i \mathbf{D}^{(i)}_k \mathbf{\Sigma}_i^T,
\end{align}
defining the distributions $p^{(i)}_k(\underline{z}) \sim \mathcal{N}(\underline{z};\underline{\mu}^{(i)}_k, \mathbf{C}^{(i)}_k)$.
The parameters $\bar{w}_k,\bar{\mu}_k$ and $\bar{\sigma}_k$ are directly estimated using the splitting library shown in Table 1. The library was generated for the standard Gaussian density (zero mean, unit variance) using the splitting technique used in \cite{hanebeck2003progressive,sankar1998experiments}. The chosen Gaussian component is first split into two mixture components starting with an initial guess 
\begin{align} \label{eq:2split}
w\exp\left\{-\frac{1}{2}\frac{(x-{\mu})^2}{\sigma^2}\right\} \approx \bar{w}_1\exp\left\{-\frac{1}{2}\frac{(x-\bar{\mu}_1)^2}{\bar{\sigma}_1^2}\right\} + \bar{w}_2\exp\left\{-\frac{1}{2}\frac{(x-\bar{\mu}_2)^2}{\bar{\sigma}_2^2}\right\},
\end{align}
where
\begin{align}
\bar{\mu}_1 = {\mu}-\epsilon,\enspace \bar{\mu}_2 = {\mu}+\epsilon, \enspace \bar{\sigma}_1=\bar{\sigma}_2=\sigma,\enspace\bar{w}_1=\bar{w}_2=w/2,
\end{align}
and $\epsilon = 0.001$ is a small constant. A least-squares error minimization is then performed wherein the standard deviation of the split components are recursively reduced while adapting the means and weights accordingly. When no further optimization is possible, the resulting two mixture densities can again be split according to (\ref{eq:2split}) yielding four mixture components.
The results of using this library for both a two-way and a four-way-split of the standard Gaussian are depicted in Fig.~\ref{fig:split}.
While maintaining a high degree of accuracy, this split is not perfect and
introduces a marginal amount of error. In theory, it would require an
infinite sum of Gaussian distributions on the r.h.s.~of (\ref{eq:2split}) to
exactly represent the single Gaussian distribution on the l.h.s.~of (\ref{eq:2split}).

\renewcommand{\arraystretch}{1.2}
\begin{table}[htbp]\centering
	\small
	\caption{Splitting library for the standard Gaussian density with zero mean and unit variance \cite{hanebeck2003progressive,sankar1998experiments}.}
	\begin{tabular}{@{}rrrr@{}}
			\toprule
			k & $\bar{w}_k$ & $\bar{\sigma}_k$ & $\bar{\mu}_k$  \\
			\midrule
			Two Way Split ($M=2$) &&& \\
			1	&	0.5000		&	0.7745125	&	-0.56520 \\
			2	&	0.5000		&	0.7745125	&	0.56520 \\
			\midrule 
			Four Way Split  ($M=4$)&&& \\
			1	&	0.127380		&	0.5175126	&	-1.41312 \\
			2	&	0.372619		&	0.5175126	&	-0.44973 \\
			3	&	0.372619		&	0.5175126	&	 0.44973 \\
			4	&	0.127380		&	0.5175126	&	 1.41312\\
			\bottomrule
	\end{tabular}
	\label{table:split}
\end{table}

\begin{figure}[htbp] 
\begin{center}
\centerline{\includegraphics[scale=0.33]{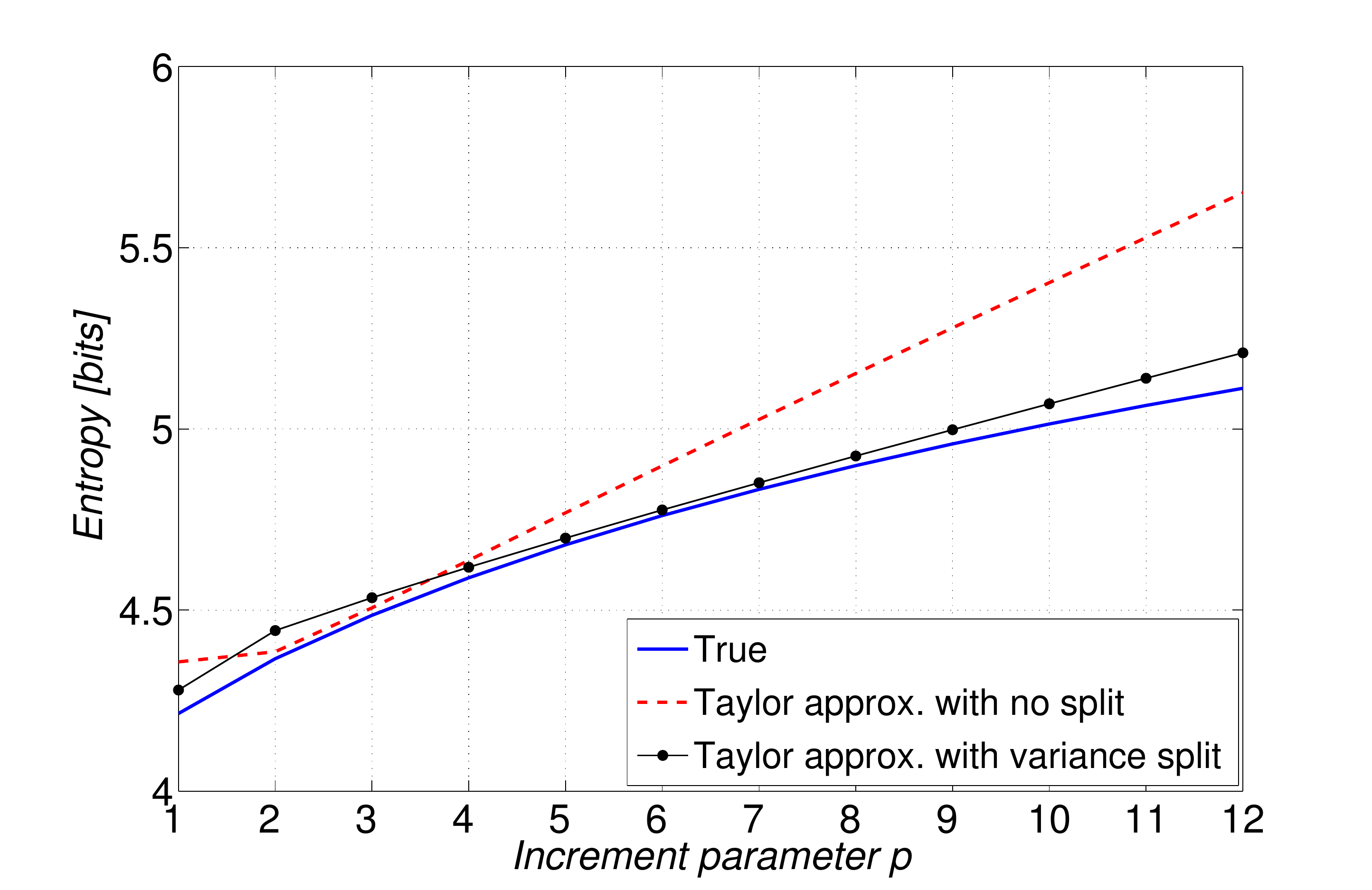}}
\caption{{Entropy approximation using a fourth-order Taylor expansion for a sample synthetic GMM with zero mean vectors, equal weights, and $\mathbf{C}_1=[4,2;2,4]$, $\mathbf{C}_2=[5+p,2;2,5+p]$, where the variance of the second distribution is incremented using the parameter \emph{p}. The true entropy was calculated using numerical integration which in this case for $n=2$ is still computationally tractable. \label{fig:comp} }}
\label{default}
\end{center}
\end{figure}

Fig.~\ref{fig:comp} exemplarily shows the simulation results for the entropy approximation of a sample GMM with $n=2$ as its variance is increased, calculated by using a fourth-order Taylor series expansion, with and without the variance split.
The results indicate an increase in the overall accuracy of the final entropy estimate when using the variance split. We have observed that the splitting approach is especially helpful at higher variances and, if required, can be performed repeatedly to further refine the approximation.  

\section{Mutual Information Results}\label{sec:4}

The MI can be trivially upper bounded as $ I(X;\underline{Y}) \leq H(X) = 1$ bit, where we have used the fact that $X$ is drawn from an equiprobable Bernoulli distribution. This upper bound depends only on the quality of the audio stimulus, independently of the subject, the audio sequence, the pre-processing of the EEG data, or the number of ROIs considered. Therefore, the maximum information that can be transferred over the ERP channel for the given input is $1$ bit.


\subsection{Single region}\label{sec:41}
We first evaluate the MI over a single region. Choosing only one ROI at a time corresponds to $\underline{Y}$ being univariate, i.e., $n=1$.
Table \ref{tab:singleROI} lists the MI results calculated for each ROI for
one test subject. Also shown are the lower and upper entropy bounds obtained using the Jacobian Log bound introduced in Section \ref{sec:331}.

The output of the ERP channel maps the activity spread over the entire cortex distributed over all 8 ROIs. 
Considering only a single region then ignores the fact that the ROIs are coupled due to the shift of neural activity between the regions. Therefore, the available information is severely limited when only a single ROI is considered. It is interesting to note however, that the temporal regions 5 and 7 show a higher MI than the other regions. This is in strong agreement with the fact that the the primary auditory cortex which is involved with the perception and processing of auditory information, is located in the temporal lobe.


\renewcommand{\arraystretch}{1.2}
\begin{table}[tbp]\centering
	\small
	\caption{Single region entropy bounds and mutual information for subject S4.}
	\begin{tabular}{@{}r|ccc|c@{}}
			\toprule
			ROI & Entropy & Lower Bound & Upper Bound & MI \\
			\midrule
			1 	& 	3.2512 	&	2.9676	&	3.5549	&	0.0464 \\
			2 	& 	3.4257 	&	3.1781	&	3.7633	&	0.0484 \\
			3 	& 	3.2650 	&	3.0493	&	3.6128	&	0.0727 \\
			4 	& 	3.0395 	&	2.7990	&	3.3828	&	0.0498 \\
			5 	& 	3.3441 	&	3.1937	&	3.7105	&	0.1481 \\
			6 	& 	3.1016 	&	2.8841	&	3.4354	&	0.0427 \\
			7 	& 	3.3437 	&	3.1089	&	3.6840	&	0.0890 \\
			8 	& 	3.5481	&	3.2253	&	3.8623	&	0.0121 \\
			\bottomrule
	\end{tabular}
	\label{tab:singleROI}
\end{table}

\subsection{Multiple regions}\label{sec:42}

\begin{figure}[htbp] 
\begin{center}
\centerline{\includegraphics[scale=0.4]{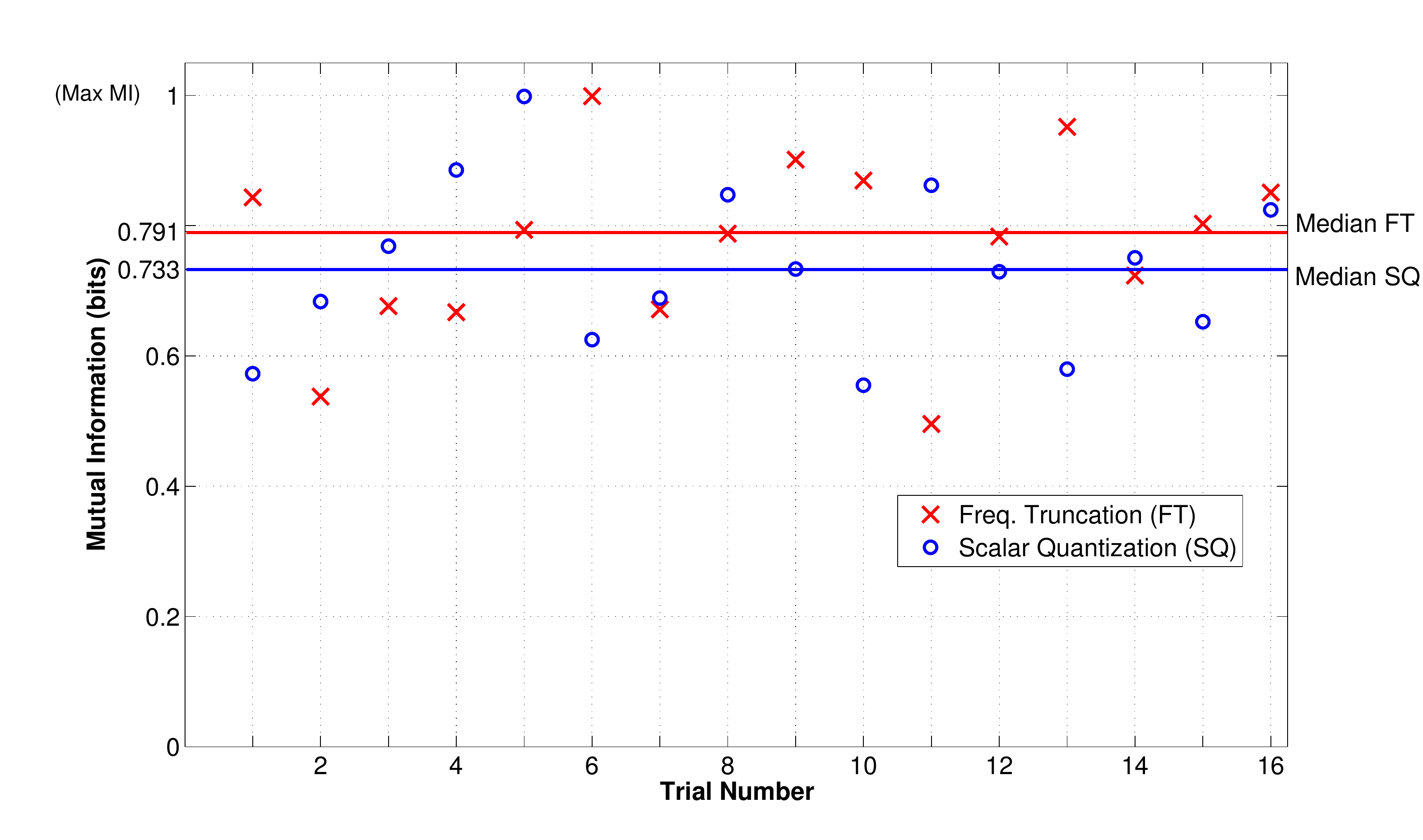}}
\caption{{MI estimates for subject S4 for the two considered distortion types, conducted using different combinations of base sequences, and time-varying distortion patterns.}}
\label{fig:mi_vs_trials_4}
\end{center}
\end{figure}

To calculate the entropy over multiple regions we use the fourth-order
Taylor series approximation presented in Section \ref{sec:332}. A
four-component variance split is performed twice to further refine the
approximation result. Fig.~\ref{fig:mi_vs_trials_4} shows the final MI
estimates of a single subject for each trial using different combinations of
the base music sequences and time-varying distortion patterns. Also,
calculated and shown in the same figure is the median MI across all trials
for each of the two distortion types. The median is chosen here to remove
any outliers among the trials. We observe that the MI over the ERP channel
for a given trial is in general moderate to high. Further, as one would expect, there is a significant increase in the MI calculated over multiple ROIs, when compared to the single ROI case. 

\begin{figure}[htpb]%
\centering
\centerline{\subfigure[Different distortion types]{\includegraphics[scale=0.42]{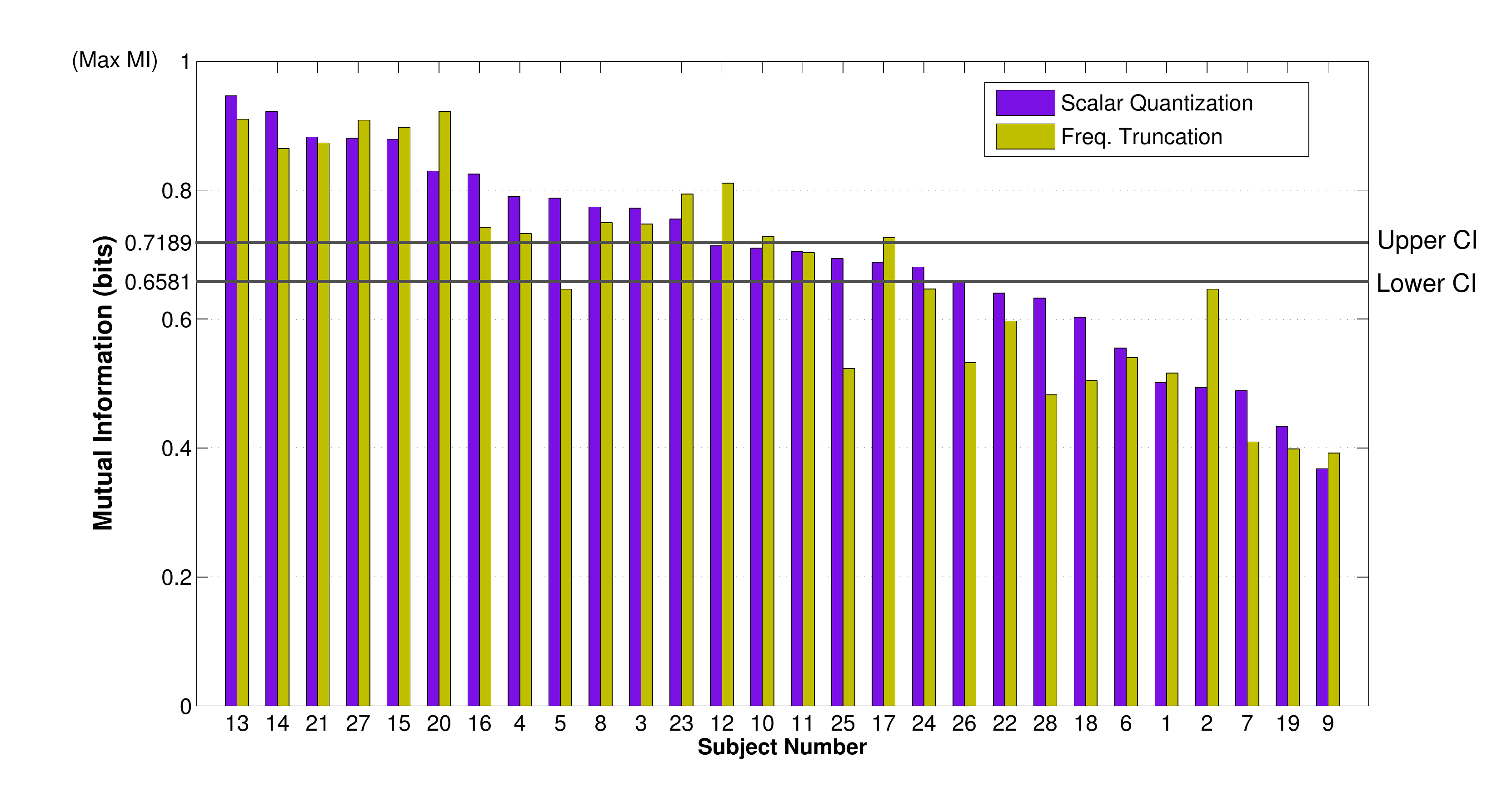} \label{fig:mivstrials}}}
\vspace{-0.3cm} 
\centerline{\subfigure[Different music types]{\includegraphics[scale=0.42]{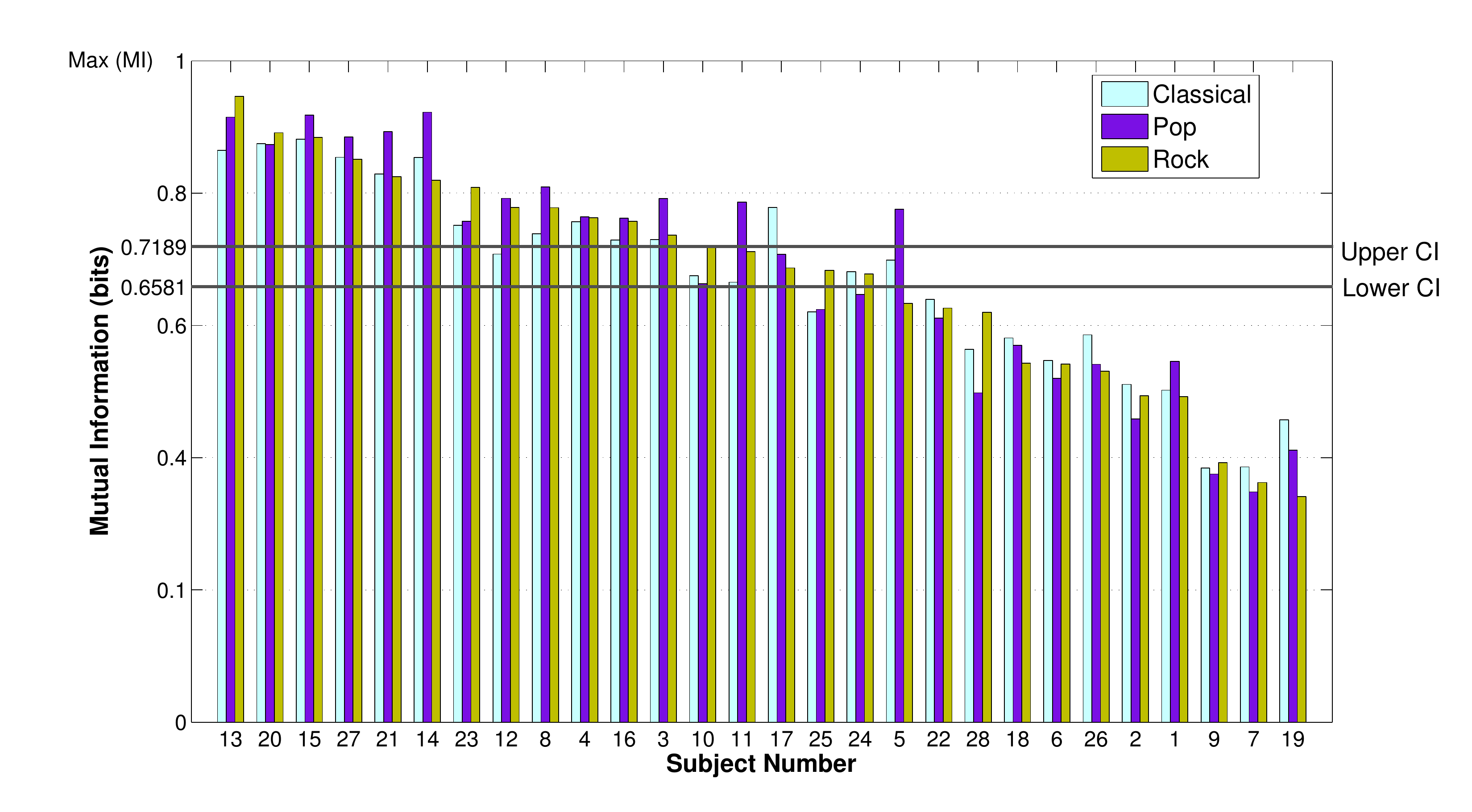}\label{fig:mivsmusic}}}
\caption{\small{Ordered median MI estimates for each of the 28 test subjects
    when presented with the same set of trial-sequences using a combination
    of different distortion types, patterns, and music sequences. Confidence
    intervals (CI) for the median are  obtained using bootstrapping \cite{efron1994introduction,efron1981nonparametric}.}}
\label{fig:mivs}
\end{figure}

The median individual MI estimates for different distortions and each subject are summarized in Fig.~\ref{fig:mivstrials}. We notice that in general for a given subject there is only a minor variation in the median MI between the different distortion types. Further, Fig.~\ref{fig:mivsmusic} shows the median MI of each subject calculated across trials for different base music sequences used. By comparing Fig.~\ref{fig:mivstrials} and Fig.~\ref{fig:mivsmusic} we observe a similar trend, and subjects that have a high MI in one experiment also have a high MI in the other. Overall, the MI is found to be consistent across the entire pool of test subjects across different trials, distortion-types, and music-types.  

\subsection{Confidence intervals for the MI using bootstrapping}\label{sec:43}
The bootstrap \cite{efron1994introduction,efron1981nonparametric} is a powerful resampling technique for assessing the accuracy of a parameter estimate (in this case the median). It is especially useful in making asymptotically-consistent statistical inferences when only a limited number of sample realizations are available. We derive confidence intervals (CI) for the median of the MI estimates using the bootstrap\textit{-t} percentile method \cite{zoubir1998bootstrap}. It has been shown \cite{politis1998computer} that the bootstrap-\textit{t} percentile intervals have a higher order accuracy as compared to regular bootstrapping by better accounting for the skewness and other errors in the original population.

The original sample set consists of the actual MI estimates calculated for each of the trials. The median of this sample is denoted by $\overline{M}$. 
Bootstrap resamples are then created by drawing samples, with replacement, from a population of this original sample set. The median for each of this resampled set is calculated and denoted as $\overline{M}^{\ast}_b$ with $b=1,2,\ldots,B$. Generally, $B$ is a large number in the order of several hundred. For our calculations here we use $B=1000$ resamples.

\km{The \textit{t-}parameter is defined by normalizing the difference between the original sample statistics (in this case the median) and the resampled statistics over the standard error}
\begin{align}
t^{\ast}_b = \frac{\overline{M}-\overline{M}^{\ast}_b}{\sigma^{\ast} / \sqrt{s}},
\end{align}
where $\sigma^{\ast2}$ is the sample variance and $s$ is the number of elements in each resampled set. The empirical standard deviation $\sigma^\ast$ is itself calculated by performing a nested bootstrap from resamples of $\overline{M}^{\ast}_b$.

\renewcommand{\arraystretch}{1.2}
\begin{table}[bt]\centering
	\small
	\caption{MI confidence intervals on  the ERP channel in bits, obtained using bootstrapping.}
	\begin{tabular}{@{}l@{\hskip 0.3cm}c@{\hskip 0.3cm}c@{}}
			\toprule
			Type & Lower CI & Upper CI \\
			\midrule
			Distortion \\
			\quad Freq. Truncation	&	0.6162	&	0.6937	\\
			\quad Scalar Quant.		&	0.6625	&	0.7087	\\
			Music \\
			\quad Rock				&	0.6594	&	0.7286	\\
			\quad Pop				&	0.6352	&	0.7321	\\
			\quad Classic			&	0.6509	&	0.7101	\\
			\addlinespace
			\textbf{Overall}		&	\textbf{0.6581}	&	\textbf{0.7189}	\\
			\bottomrule
	\end{tabular}
	\label{tab:boots}
\end{table}

\km{We therefore get a total of $B$ values of the \textit{t-}parameter $(t^{\ast}_1,\ldots,t^{\ast}_b,\ldots,t^{\ast}_B)$, from which we calculate the required percentile cutoffs of our CIs. For our analysis, we take the  2.5th and 97.5th percentile cutoffs and calculate the lower and upper CI, respectively, as}
\begin{align}
\mbox{CI$_{\text{low}}$} &= \overline{M}-t_{25}\sigma^\ast , \notag \\
\mbox{CI$_{\text{high}}$} &= \overline{M}-t_{975}\sigma^\ast .
\end{align}
Table \ref{tab:boots} lists the confidence intervals for different distortion and music types. Both upper and lower CIs are also included in Fig.~\ref{fig:mivs}.



\section{Discussion}
\km{MI quantifies the amount of information the recorded EEG contains about
  the quality of the input audio-stimulus. By measuring how aligned the brain activity is in response to the stimulus we thereby directly quantify the subject's perception  with respect to the change in audio quality with time. A high MI therefore indicates a large correlation
between the actual and the perceived audio quality. For example, the fact that subjects 13 and 27 have a high median MI in Fig.~\ref{fig:mivs} implies that these subjects have a strong perception towards quality change irrespective of the music  or distortion type used in the test sequence.} 

\jkl{However, brain dynamics are fundamentally multiscale in nature and it is
  often difficult to draw a one-to-one relation between the neural
  electrical field dynamics and the cognitive/behavioral state
  \cite{makeig2012evolving}. The EEG activity is dependent on the
  psychological state of the subject including concentration, motivation,
  and fatigue. 
As a consequence, the recorded EEG activity varies considerably not only
over the duration of a trial, but also from one trial to another, which
could result in low MI, for example as it is the case with subjects 7, 9, and
19.} 

 
It is also important to note the significance of choosing the ROIs. The
output EEG activity recorded over the ERP channel is dependent on the number
of regions, the size (i.e., the number of individual electrodes within each
ROI), and the location of the ROIs on the scalp. Increasing the number of
regions increases complexity whereas considering too few oversimplifies the
problem, as demonstrated by the single ROI results in
Table~\ref{tab:singleROI}. In our work we select the regions based on the
cortical lobes (see Section \ref{sec:ROI}). \jkl{The ROI selection could possibly
be improved by identifying the distribution of virtual sources inside the
brain which generate the EEG activity on the scalp.} A source localization of the EEG signal is commonly referred to as the \textit{inverse problem} and while several solutions have been suggested to varying degrees of accuracy \cite{grech2008review} it continues to be an undetermined problem \cite{makeig2012evolving}.



\section{Conclusion}


In this work we have presented a new framework for
audio quality assessment based on observed EEG data for subjects
listening to time-varying distorted audio. 
 By modeling the end-to-end perceptual processing chain as a discrete-input
 time-varying nonlinear SIMO channel with memory, 
we are able  to quantify how the subjective perception changes if the observed
  audio quality changes with time. We have focused on the simplest
binary-input SIMO case where the audio quality can only change between distorted
and undistorted levels. The MI results demonstrate the practical applicability of
this approach and provide a performance measure for the
information transfer over the ERP channel, which can be computed via a new
low complexity algorithm. 
This algorithm is based on a fast approximation technique for the differential entropy
of a multidimensional GMM. The proposed approach can be extended to video
quality perception and to other neuroimaging assessment techniques like MEG,
albeit with a higher complexity. As potential future work the proposed information theoretic
framework can be exploited to develop an inference model for blind audio quality classification based on perception. 



\appendix
In this appendix we derive the Taylor series expansion of the entropy up to its fourth order component. The Taylor series expansion of the entropy for the GMM can be written as
\begin{align}
	h(\underline{z})  &= -\sum\limits_{i=1}^{L}\int\limits_{\mathbb{R}^n}\!\!w_i p_i(\underline{z})\cdot \left\{f(\underline{\mu}_i) + \frac{f^{''}(\underline{\mu}_i)}{2!}{(\underline{z}-\underline{\mu}_i)}^2 + \ldots\right\} d\underline{z} \notag \\
				& = h_0 + h_2 + h_4 + R,							
\end{align}
where $h_0, h_2, h_4$ are the zeroth, second and fourth order expansion terms respectively, and $R$ is the residual sum term representing all the remaining higher order terms.
\subsection{Zeroth-order term}
The zeroth-order Taylor series expansion term $h_0$ is given by
\begin{align}  \label{taylor5}
	h_0 &= -\sum\limits_{i=1}^{L}\int\limits_{\Re^n}\!\!w_i
        p_i(\underline{z}) f(\underline{\mu}_i) \, d\underline{z} \notag \\
        &= -\sum\limits_{i=1}^{L} w_i\log p(\underline{\mu_i}).
\end{align}
which follows from the fact that the integral over the entire probability distribution is 1.
\subsection{Second-order term}
The second-order expansion term requires us to calculate the gradient $\mathsf{g}_i$ and Hessian $\mathsf{H}_i$ of the Gaussian probability distribution $p_i(\underline{z})\sim\mathcal{N}(\underline{z};\underline{\mu}_i,\mathbf{C}_i)$, which are respectively defined as
\begin{align}  \label{taylor6}
	\mathsf{g}_i	&\triangleq \: \nabla p_i(\underline{z}) \:=\: p_i(\underline{z})\mathbf{C}_i^{-1}(\underline{\mu}_i - \underline{z}), \\
	\mathsf{H}_i	&\triangleq \: (\nabla\nabla^{T}) p_i(\underline{z}) \:=\: \frac{\mathsf{g}\mathsf{g}^T}{p_i(\underline{z})}-p_i(\underline{z})\mathbf{C}^{-1},
\end{align}
where $\nabla$ denotes the vector differential operator and $i=1,2,\dots,L$. Accordingly, the gradient $\mathfrak{g}_i$ and Hessian $\mathbf{H}_i$ for the log-density function $f(\underline{z})$ can be defined as
\begin{align}  \label{taylor7}
	\mathfrak{g}_i	\:&\triangleq\: \nabla f(\underline{z}) \:=\: \nabla \log{p_i(\underline{z})} \:=\: \frac{1}{p_i(\underline{z})}\mathsf{g}_i, \\
	\mathbf{H}_i	&\triangleq (\nabla\nabla^{T}) f(\underline{z}) = -\frac{1}{p_i(\underline{z})^2}\mathsf{g}_i\mathsf{g}_i^T + \frac{1}{p(\underline{z})}\mathsf{H}_i.
\end{align} 
We can then calculate the second-order term as
\begin{align}  \label{taylor8}
	h_2	&=  -\sum\limits_{i=1}^{L}\int\limits_{\mathbb{R}^n}\!\!w_i p_i(\underline{z})\cdot \frac{1}{2}\mathbf{H}_i\cdot (\underline{z}-\underline{\mu}_i)(\underline{z}-\underline{\mu}_i)^T \,d\underline{z}\notag \\
			 &= -\frac{1}{2}\sum\limits_{i=1}^{L} w_i \cdot \mathbf{H}_i \circ \mathbf{C}_{i}\Bigl{\rvert}_{\underline{z}=\underline{\mu}_i}\quad.
\end{align}
The symbol $\circ$ denotes the Frobenius product $\mathbf{A}\circ\mathbf{B} = \sum\limits_{i} \sum\limits_{j}a_{ij}\cdot b_{ij}$.
\subsection{Fourth order term}
The third order partial derivative with respect to $p_i(\underline{z})$,
$i=1,2,\dots,L$, is given by
\begin{align}
	\mathsf{T}_i  \:&\triangleq\: \mathsf{H}_i\!\otimes\!\left(\mathbf{C}_i^{-1}(\underline{z}-\underline{\mu}_i)\right) - \mathsf{g}_i\!\otimes\!\mathbf{C}_i^{-1} - \#\mathbf{C}_i^{-1}\mathsf{g}_i^{T},
\end{align}
where $\mathbf{A}\!\otimes\!\mathbf{B}$ is the Kronecker product of the matrices $\mathbf{A},\mathbf{B}$, and $\#$ is the matrix vectorization operator. The vectorization of a matrix is the linear transformation of converting a $m\times n$  matrix into a $mn\times 1$ column vector by stacking the columns of the matrix on top of each other. Further, partial differentiating the Hessian twice yields the fourth order derivative with respect to $p_i(\underline{z})$,
\begin{align}
\mathsf{F}_i \:&\triangleq\: (\nabla\nabla^{T})\mathsf{H}_i \notag \\
 &= \mathsf{H}_i\!\otimes\!\mathbf{C}_i^{-1} + \mathbf{C}_i^{-1}(\underline{z}-\underline{\mu}_i)\!\otimes\!\mathsf{T}_i^T + \mathbf{C}_i^{-1}\!\otimes\!\mathsf{H}_i + \mathbf{C}_i^{-1}\!\otimes\!\mathsf{H}_i^{T}.
\end{align}
Similarly then, the third and fourth order partial differential matrix for the log-density $f(\underline{z})$ are respectively defined as
\begin{align}
	\mathbf{T}_i  \:&\triangleq\: \frac{1}{p_i(\underline{z})^3}\left\{\mathsf{g}_i\!\otimes\!\mathsf{g}_i\mathsf{g}_i^T \right\} \notag \\
	& \qquad -\frac{1}{p_i(\underline{z})^2} \left\{  \mathsf{H}_i\!\otimes\!\mathsf{g}_i + \mathsf{g}_i\!\otimes\!\mathsf{H}_i  \right\} \notag \\
	& \qquad\qquad + \frac{1}{p_i(\underline{z})} \left\{ \mathsf{T}_i - \#\mathbf{C}_i^{-1}\mathsf{g}_i^T \right\}, \\
	\mathbf{F}_i  \:&\triangleq\: \frac{1}{p_i(\underline{z})^4}\left\{\mathsf{g}_i^T\mathsf{g}_i\!\otimes\!\mathsf{g}_i\mathsf{g}_i^T \right\} \notag \\
	&\quad -\frac{1}{p_i(\underline{z})^3}\left\{ \mathsf{H}_i\!\otimes\!\mathsf{g}_i\mathsf{g}_i^T + \mathsf{g}_i\!\otimes\!\mathsf{H}_i\!\otimes\!\mathsf{g}_i^T		\right\} \notag \\
	&\qquad +\frac{1}{p_i(\underline{z})^2}\left\{ 2\mathsf{g}_i\!\otimes\!\mathsf{T}_i^T - (\#\mathbf{C}_i^{-1}\mathsf{g}_i^T)\!\otimes\!\mathsf{g}_i^T\right\} \notag \\
	&\qquad\quad +\frac{1}{p_i(\underline{z})} \left\{ \mathsf{F}_i -2\mathsf{H}_i\!\otimes\!\mathbf{C}_i^{-1} - \mathsf{T}_i\!\otimes\!\mathsf{g}_i^T + \mathbf{C}_i^{-1}\!\otimes\!\mathsf{H}_i^T	\right\}.
\end{align}
The fourth order term is then given by 
\begin{align}  \label{taylor8}
	h_4	&=  -\sum\limits_{i=1}^{L}\int\limits_{\mathbb{R}^n}\!\!w_i p_i(\underline{z})\frac{f^{''''}(\underline{\mu}_i)}{4!}{(\underline{z}-\underline{\mu}_i)}^4\,d\underline{z}\notag \\
			 &= -\frac{1}{24}\sum\limits_{i=1}^{L} w_i \cdot \mathbf{F}_i \circ (\mathbf{C}_i\otimes\mathbf{C}_i^T)\Bigl{\rvert}_{\underline{z}=\underline{\mu}_i}\quad.
\end{align}

\vspace{5mm}

\bibliographystyle{IEEEtran}
\bibliography{refs}

\end{document}